\documentclass{emulateapj}
\newcommand{\HI}{\ion{H}{1}~}

\newcommand{\kms}{km~s$^{-1}$ }
\newcommand{\Lya}{Lyman~$\alpha$~}

\usepackage{lscape} 
\usepackage{color}

\slugcomment{Received: April 27, 2016; Accepted: July 13, 2016}

\shorttitle{Distribution of Cold Gas in the Inner Halos of Galaxies}
\shortauthors{Borthakur}

\begin{document}

\title{Distribution of Cold ($\lesssim 300$~K) Atomic Gas in Galaxies: Results from the GBT\altaffilmark{1} HI Absorption Survey Probing the Inner Halos ($\rho<20$~kpc) of Low-z Galaxies}

\altaffiltext{1}{Based on observations with the Robert C. Bryd Green Bank Telescope of the
National Radio Astronomy Observatory, a facility of the National
Science Foundation operated under cooperative agreement by Associated
Universities, Inc.}

\author{Sanchayeeta Borthakur}
\affil{Department of Physics \& Astronomy, Johns Hopkins University, Baltimore, MD, 21218, USA and The Astronomy Department, University of California, Berkeley, CA USA }
\email{sanch@jhu.edu}

\begin{abstract}
We present the Green Bank Telescope absorption survey of cold atomic hydrogen ($\lesssim 300$~K) in the inner halo of low-redshift galaxies. The survey aims to characterize the cold gas distribution and to address where condensation - the process where ionized gas accreted by galaxies condenses into cold gas within the disks of galaxies - occurs.  
Our sample consists of 16 galaxy-quasar pairs with impact parameters of $\le$ 20~kpc. 
We detected an \HI absorber associated with J0958+3222 (NGC~3067) and \HI emission from six galaxies. We also found two \ion{Ca}{2} absorption system in the archival SDSS data associated with galaxies J0958+3222 and J1228+3706, , although the sample was not selected based on the presence of metal lines. Our detection rate of \HI absorbers with optical depths of $\ge 0.06$ is $\sim$7\%. We also find that cold \HI phase ($\lesssim$~300~K) is 44($\pm$18)\% of the total atomic gas in the sightline probing J0958+3222. We find no correlation between the peak optical depth and impact parameter or stellar and \HI radii normalized impact parameters, $\rho/\rm R_{90}$ and $\rho/\rm R_{HI}$. We conclude that the process of condensation of inflowing gas into cold ($\lesssim$ 300~K) \HI occurs at the $\rho << 20$~kpc. However, the warmer phase of neutral gas (T $\sim$ 1000~K) can exists out to much larger distances as seen in emission maps. Therefore, the process of condensation of warm to cold \HI is likely occurring in stages from ionized to warm \HI in the inner halo and then to cold \HI very close to the galaxy disk.

\end{abstract}


\section{INTRODUCTION\label{intro}}

Theoretical and observational evidence indicate that gas inflow is essential to the growth of galaxies. The observed star-formation histories and stellar population metallicities (e.g., the G-dwarf problem) require continuous accretion of low-metallicity gas into galaxies in the present epoch. Simulations by \citet[][and others]{birn03,keres05} have shown how gas ($\rm 10^4~K$) may be accreted by galaxies via filamentary structures from the cosmic web. 
Hence, understanding the connection between cold atomic hydrogen (at $\rm \approx 10^2~K$) within the disks, the hot halo gas (at $\rm \approx 10^6~K$), and the cooler accreting gas (at $\rm T \approx 10^{4-5}~K$) is essential to our understanding of galaxy growth and evolution.
On the observational front, cool gas traced by Lyman~$\alpha$ absorbers has been ubiquitously detected in the circumgalactic medium (CGM) of galaxies. \citet{werk14}  found that the total mass in the CGM can be as large as the stellar mass of a galaxy. The strength of the CGM absorbers increases as we probe closer to the galaxies \citep[][and references therein]{lanzetta95, chen98, tripp98, chen01b, bowen02, prochaska11, stocke13, tumlinson13, liang14, borthakur15}, thus suggesting the possibility of an active condensation process in the inner regions of a galaxy halo. Recently, \citet{curran16} observed a similar inverse correlation between 21cm \HI absorption strength and impact parameter by combining data from various \HI absorption surveys. Furthermore, \citet[COS-GASS survey][]{borthakur15} found that the strength of the \Lya absorbers in the CGM is strongly correlated with the amount of atomic hydrogen within the inner regions of the galaxies.

While there is growing observational indications that the inflowing/accreting gas eventually condenses into the \HI disk, the details of the process of condensation has eluded us so far. There is very little direct observational evidence on how the cool accreting gas condenses into neutral \HI\ and descends into the disks of galaxies. 
Part of the problem is that condensing cold ($\rm 10^{2-4}~K$) clouds are expected to have low column densities, which posses a serious limitation in detecting them around distant galaxies. 
Deep \HI\ imaging studies of Milky Way shown the presence of extra-planar gas \citep{kalberla08} that exists as filaments extending up to several kiloparsecs (Hartmann 1997). 
The extraplanar gas amounts to 10\% of the total \HI\ in our galaxy \citep{kalberla08}.
Recently, there is growing evidence of the presence of extraplanar cold gas in a few nearby galaxies where \HI imaging of the faint low column density gas is possible have shown the ubiquitous presence of extra-planar gas \citep[][HALOGAS survey]{oosterloo07,heald11}. 
\HI imaging of NGC~891 by \citet{oosterloo07} revealed a large extended gaseous halo that contain almost 30\% of its total \HI\ in the form of \HI\ filaments extending up to 22~kpc vertically from the galactic disk. 
While the origin of such structures is not clear, possible origins include \HI\ accreted via satellite galaxies and/or from the inter-galactic medium or condensation of hot halo gas into \HI\ structures.  
Besides extraplanar gas, the halo of Milky Way contains high and intermediate velocity clouds (HVCs and IVCs respectively) with velocities deviating from the \HI\ in the disk with a covering fraction of $\sim$ 37\% at log~N(HI)$> 17.9$ \citep{lehner12}. Another population of low-mass \HI\ clouds in the galactic halo with peak column densities of 10$^{19} \rm cm^{-2}$, and sizes of a few tens of parsecs exists in the Milky Way halo \citep{lockman02}.
The existence of these extraplanar clouds in the Milky Way has been known for decades now, nevertheless their origin still remains controversial. 
One likely scenario is that these \HI structures are tracing the gas transport route from the outer CGM (50-200 kpc) to the inner regions (5-10kpc) of a galaxy or vice-versa.

 \begin{figure*}
\includegraphics[trim = 55mm 20mm 55mm 20mm, clip,scale=1.07,angle=-0]{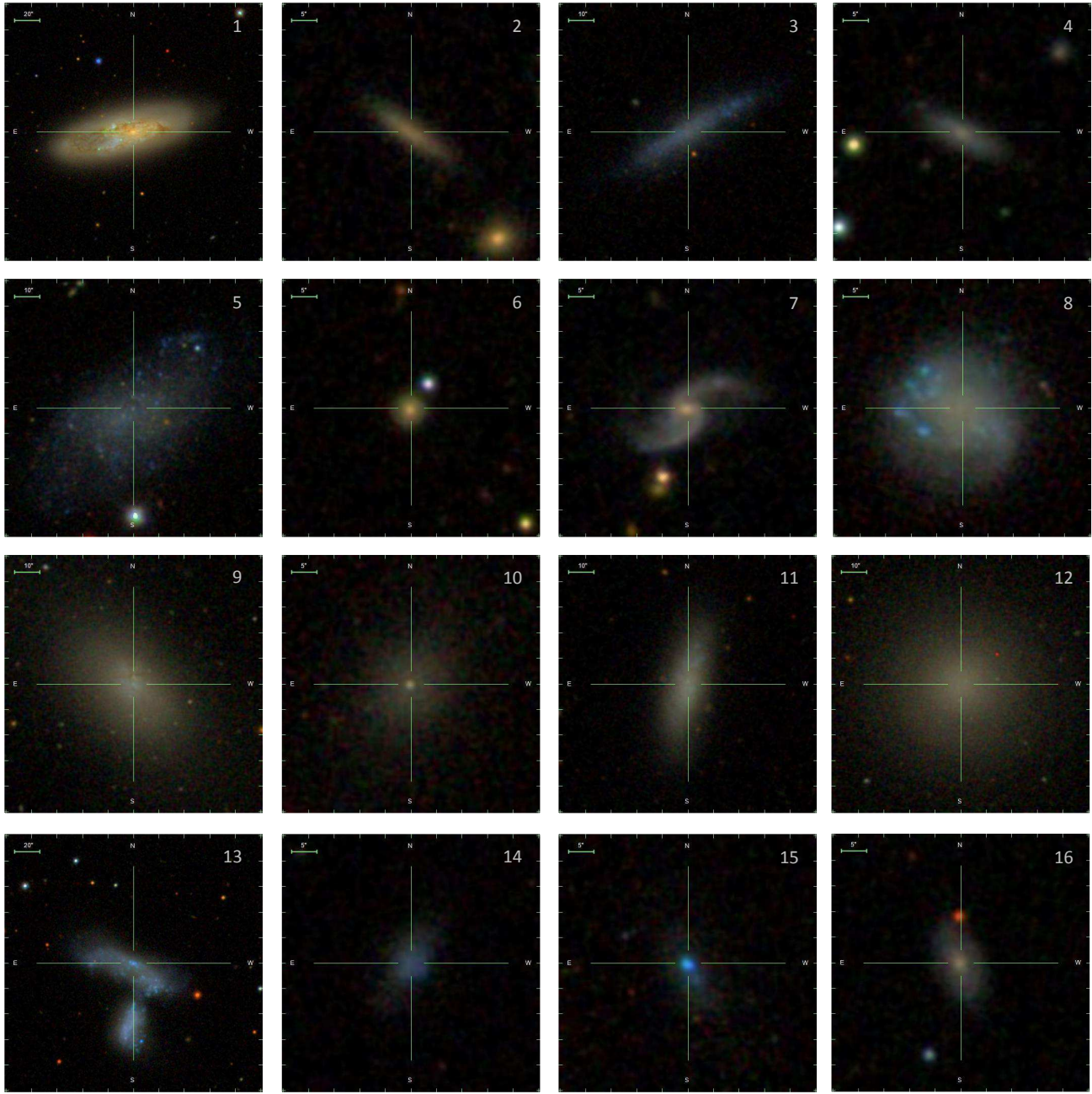}    
\caption{Postage stamp images showing SDSS false color images of galaxies from our sample. The cross-bar is centered at the target (foreground) galaxy. The target ID is printed in the top right corner of the image. Further details on the targets are presented in Table~1.  }
 \label{fig-SDSS_images} 
\end{figure*}

 \begin{figure*}
\includegraphics[trim = 20mm 0mm 54mm 0mm, clip,scale=0.60,angle=-0]{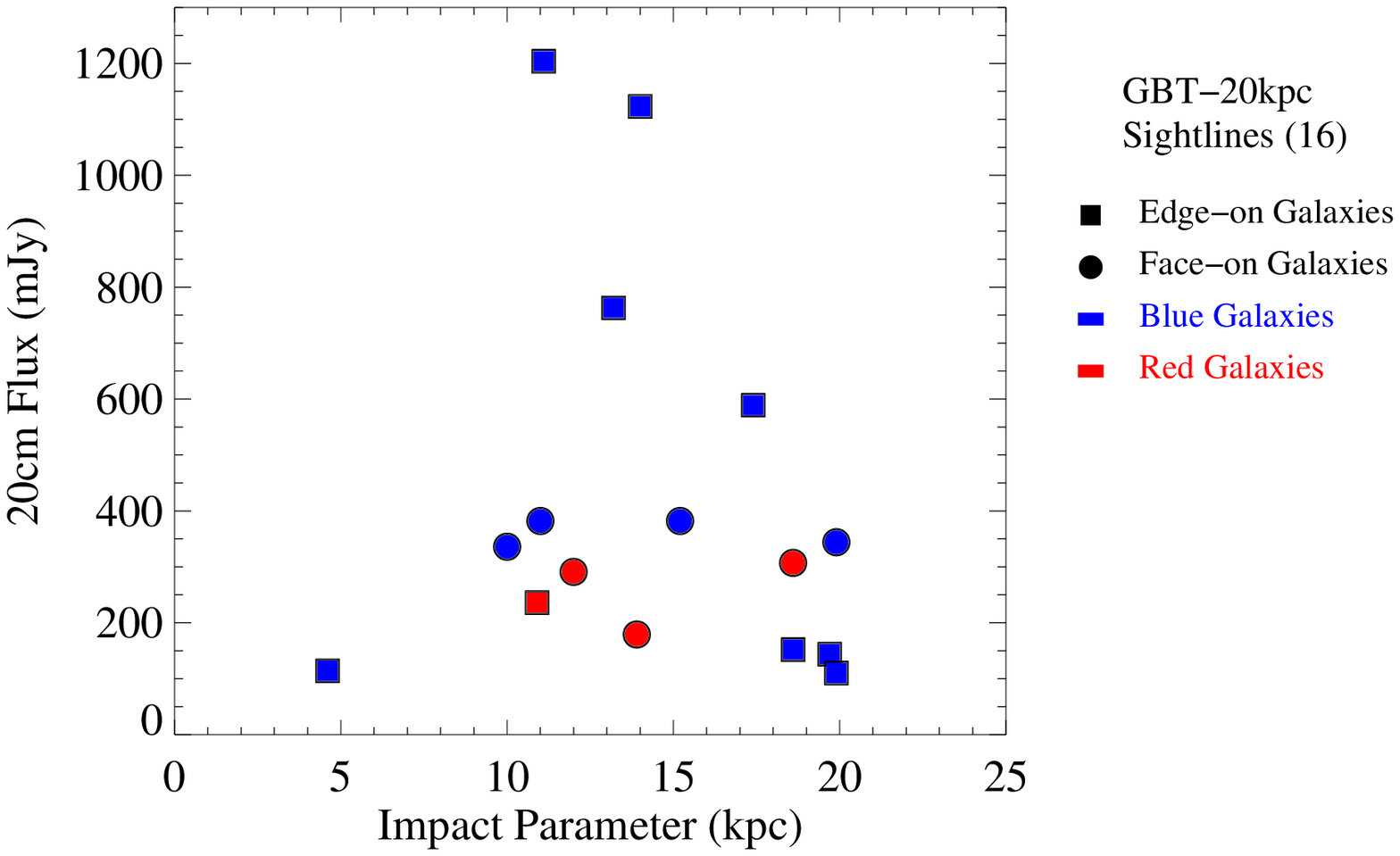}   
\includegraphics[trim = 10mm 0mm   0mm 0mm, clip,scale=0.60,angle=-0]{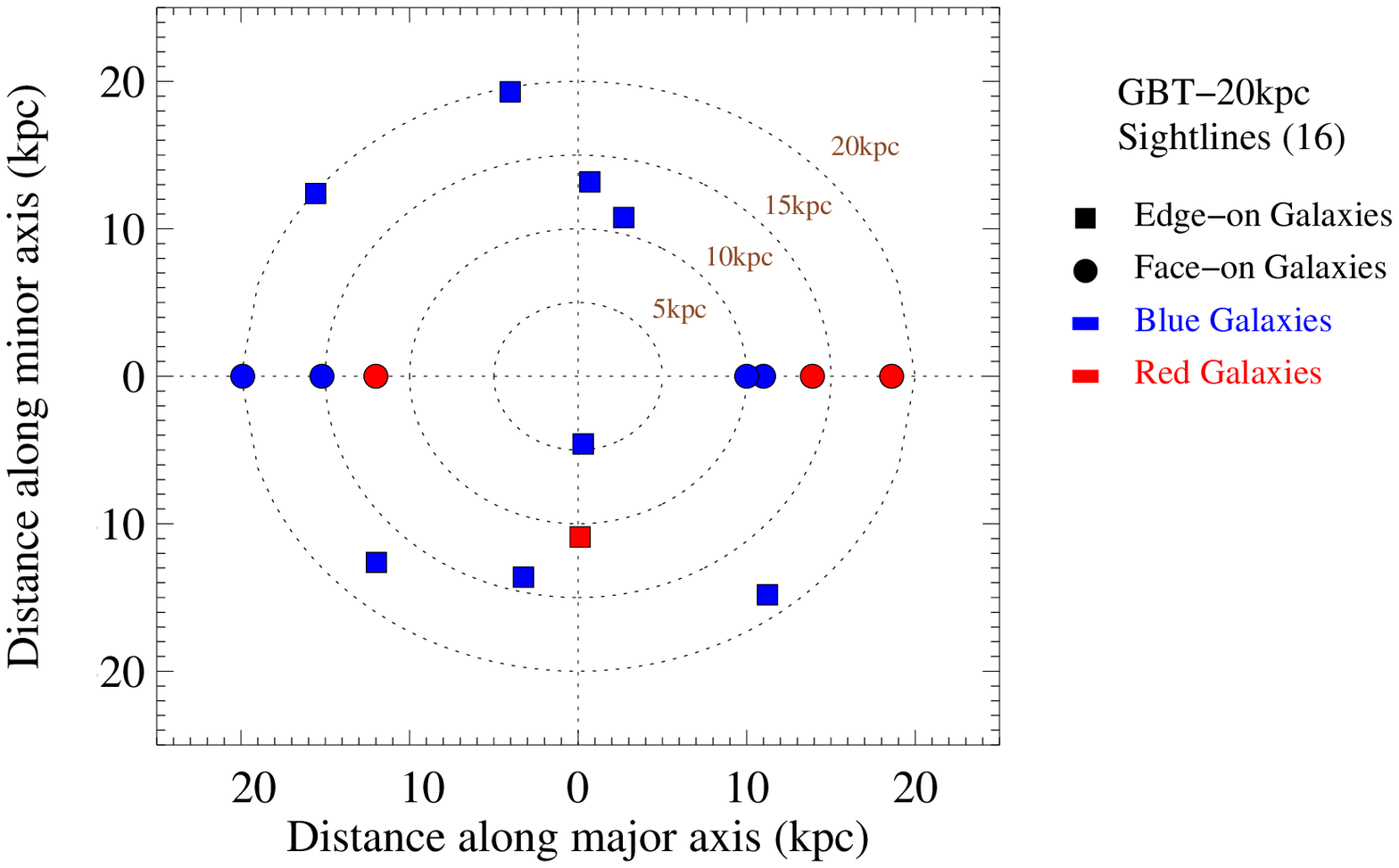}    
\caption{Left: Impact parameter of the sightline to the foreground galaxy is plotted against the 20cm radio continuum flux of the background source for our sample of 16 galaxy-quasar pairs. The galaxies are marked with different color indicating the galaxy ``color" (a proxy for the star-formation) and  symbols indicating the orientation of the foreground galaxy with respect to the sightline. Right: The position of the sightline with respect to the major and minor axis of the foreground galaxy. The sightlines that probe face-on galaxies are plotted along the minor axis as filled circles.  }
\label{fig-rho-S20}
\end{figure*}

 While the present-day instruments are able to achieve sufficient surface brightness sensitivity to detect these clouds in emission around nearby galaxies (e.g. 240~hrs of WSRT for NGC~891 to get $\rm N(HI)= 1 \times 10^{19} ~cm^{-2}$ with at a spatial resolution of 23.4$^{\prime\prime}~\times \rm 16.0^{\prime\prime}$ and spectral resolution of 8~\kms by \citet{oosterloo07}), similar experiments for the more distant galaxies are  extremely challenging.
A solution to this problem is to conduct a statistical study using Quasar absorption line spectroscopy. 
HI associated with the galaxies may be probed via corresponding absorption features in the spectrum of background Quasars.
There are two main advantages of this method - (1) it allows us to probe \HI at column densities orders-of-magnitude lower than that in emission studies; and 
(2) the strength of detection does not depend on the redshift of the galaxy, thus making such a study unbiased for redshift related affects. 

A large body of work over the past several decades have taken advantage of radio bright background sources to probe \HI properties of foreground galaxies \citep[][and reference therein]{haschick75, rubin82, haschick83, briggs83, boisse88, corbelli90, carilli92, kanekar01, kanekar02, hwang04, keeney05, borthakur10b,  gupta10, borthakur11, srianand13, gupta13, borthakur14b, reeves15, zwaan15, dutta16, reeves16, curran16}. Most of these studies targeted lower redshift galaxies as the galaxy redshift catalogs from which the sample were generated become increasingly sparse at higher redshifts. Some of these studies, such as by \citet[][G10 hereafter]{gupta10}  and \citet[][B11 hereafter]{borthakur11}, have been surveys that have explored the covering fraction of cold \HI in the halos of the foreground galaxies. G10 found the covering fraction of \HI\ to be $\sim$50\% within 15~kpc  of any galaxies. B11 too found the same and concluded that covering fraction falls off rapidly with distance from the center of the galaxy. One caveat of these studies is the limited and somewhat biased samples. For a full census of cold gas around galaxies, one would require an unbiased sample as well as sensitive observations with high spatial and spectral resolution. 

Here we present our study utilizing the highly sensitive data from the Robert C. Byrd Green Bank Telescope (GBT). Detailed descriptions of our sample, the GBT observations and data reduction are presented in Section~2.  The results are presented in  in Section~3 and their implications are discussed in Section~4. Finally, we summarize our findings in Section~5. The cosmological parameters used in this study are $H_0 =70~{\rm km~s}^{-1}~{\rm Mpc}^{-1}$, $\Omega_m = 0.3$, and $\Omega_{\Lambda} = 0.7$.

\section{OBSERVATIONS  \label{sec:observations}}

\subsection{Sample \label{sec:sample}}

In order to fill the void in our current understanding and to have an unbiased census of \HI\ distribution around galaxies, we conducted the GBT observations of a complete radio selected sample of background Quasars from the VLA FIRST Survey and foreground galaxy pairs selected and Sloan Digital Sky Survey (SDSS). This survey aims to characterize the cold ($\lesssim300$~K) neutral gas distribution within 20~kpc of the galaxies. 
Vast amount of ancillary data for the galaxies including photometry and spectra are already available from the SDSS. For example, galaxy properties like the star formation rates and orientation \footnote{Based on position angle from SDSS photometry \citep[see][for details]{borthakur13}} are obtained from the SDSS and SDSS based catalogs. Multicolor SDSS images of the target galaxies are presented in Figure~\ref{fig-SDSS_images}.

The sample was assembled by cross correlating radio quasars from the VLA FIRST and SDSS based Quasar catalogs by \citet{richards09} with the 7th data release of the SDSS spectroscopic galaxy survey \citep{abazajian09}. This yielded a sample of 16 Quasar-galaxy pairs with background Quasar flux at 20~cm, $\rm S_{Quasar} \ge $~100~mJy and impact parameter, $\rm  \rho\le 20$~kpc (see left panel of Figure~\ref{fig-rho-S20}). The sample also covers a wide range of orientations of the Quasar sightline with respect to the major axis of the galaxy (see right panel of Figure~\ref{fig-rho-S20}).
  The sample covers galaxies with luminosity ranging from 0.0001 to 1.5~L$_{*}$ within redshift range, 0.001 $\le$ z $\le$ 0.18. Details of the sample are presented in Table~1.

This is for the first time that a complete SDSS based sample has been selected for 21~cm study solely based on the 20~cm flux of the background Quasars and impact parameter to the foreground galaxy. A similar sample of southern galaxies was studied by \citet{reeves15, reeves16}. Previous SDSS-based studies like G10 and B11 selected their sample based on either optical properties of the QSO. In fact, G10 and \citet{gupta12} also included sightlines that were known to exhibit \ion{Ca}{2} and \ion{Na}{2} absorbers. On the other hand, B11 used optically bright spectroscopically confirmed QSOs that have radio counterpart as their main criterion and had missed multiple radio bright sources that didn't have bright optical counterpart. Hence, our sample is unbiased in terms of the optical brightness of the background source or the presence of metal-lines. Our sample does include a few targets that have been observed before. For uniformity of the observations and data reduction, we re-observe them in the same setting as the remaining targets so that observational effects are minimized. Also it is worth noting that our sample do not probe \HI in self-absorption associated with radio galaxies \citep[unlike recent studies by ][and references therein]{allison14, gereb14}, which may have very different ISM conditions (especially near the radio engine) than normal galaxies.

\subsection{GBT Observations}

We observed 16 Quasar-Galaxy pairs with the GBT in L-band for a total 34~hours under observing program GBT-14B-024 and GBT-15B-363.  We used the dual polarization L-band system using backend VEGAS. We employed two intermediate frequencies (IFs) each with a total bandwidth 11.7~MHz spread over 32768 channels centered at  1420.405~GHz, 1667.358~GHz respectively. These corresponds to rest-frequency for \HI 21~cm and OH maser lines. This setup provided us with a channel width of 0.35 kHz corresponding to 0.075~\kms.

 We spent about 30~mins, 1~hr,  and 4~hrs on-source for targets probed by background Quasar with flux at 1.4~GHz $\rm S_{1.4GHz}>300~mJy$ (9~targets), $\rm 300< S_{1.4GHz}<200~mJy$ (2~targets), and $\rm 200<S_{1.4GHz}<100~mJy$ (5~targets), respectively. This allowed us to achieve limiting optical depth of $\rm \tau_{3\sigma}\le 0.06$ across the sample despite the order-of-magnitude variation in the flux of the background source.
 The limiting optical depth translates to a limiting column density  ($\rm \Delta V=1~kms^{-1}$) to N(HI)= 1.823~$\rm \times~10^{18}~\frac{T_s}{\it f}~\int \tau {\it dv} ~cm^{-2}~\equiv $ 3~$\times~\rm 10^{16}~\frac{T_s}{\it f}$, where $\rm T_s$ is the spin temperature and $f$ is the covering fraction of the background source. 
 If $\rm T_s= 100~K$ and $f=1$, then N(HI) $\sim$ 3~$\times~\rm 10^{18} ~cm^{-2}$, which is more than 3 times more sensitive than that achieved for NGC~891 by \citet{oosterloo07} for a fraction of the observing time and with much higher spectral resolution. 
 
 The excellent sensitivity of the GBT makes it an ideal instrument to carry out observations of faint gas in absorption. The high spectral resolution provided by the GBT makes it possible to detect extremely cool components ($\rm T_s\sim3-300~K$) that might have line widths of less than a \kms. However, one disadvantage of the GBT is its large beam  and as a result of which \HI absorption may be filled in with \HI emission from another region of the host galaxy. For our sample, we do not expect this to affect the sight lines where the Quasars are brighter than 200~mJy (or integration time for less than 1~hour). The spectra for these sightlines have significant low signal-to-noise (S/N) that hinders the detection of \HI in emission. At the same time these sightlines should show larger depth of the 21~cm absorption feature  (for the a fixed optical depth) due to the brightness of the background Quasar.  Only for the sightlines where the background Quasars are faint and we integrate for 4~hrs have good S/N to detect 21~cm \HI emission that might be filling in the weak absorption. In such cases, the GBT observations act as pilot observations and we plan to follow these targets with the Karl G. Jansky Very Large Array (VLA) that would allow us to probe these sightlines with high spatial resolution mitigating the effect of emission filling in possible absorption. We summarized the observational parameter in Table~2.

 \subsection{Data Reduction \label{sec:observations_datareduction}} 
 
 The data were analyzed using the interactive software package GBTIDL that was specifically designed for reduction and analysis of spectral line data obtained with the GBT. We obtained composite spectra by combining radio frequency interference (RFI)-free integrations using GBTIDL routines. The co-added spectra were fitted with a low-order polynomial (second or third order in most cases) for identifying the baseline. 
The good bandwidth stability of VEGAS has allowed us to fit a lower order polynomial, therefore, making our data sensitive to broad spectral features that may be associated with \HI emission from the host galaxy. In addition we chose a large range in velocity to fit for the baseline and hence we expect to detect individual emission features that are narrower than $\sim$1500~\kms.

 \begin{figure}
\includegraphics[trim = 0mm 0mm 0mm 0mm, clip,scale=0.65,angle=-0]{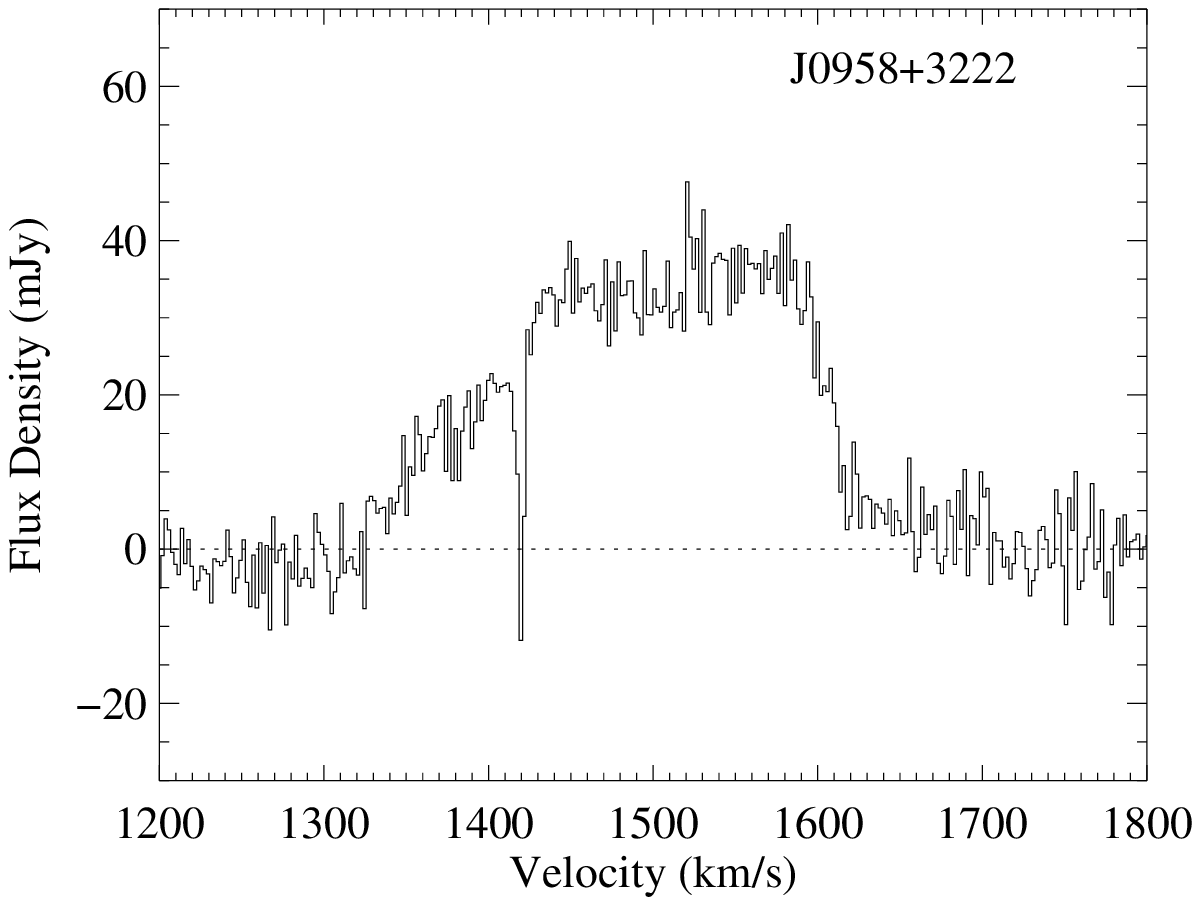}   
\includegraphics[trim = 0mm 0mm 0mm 0mm, clip,scale=0.65,angle=-0]{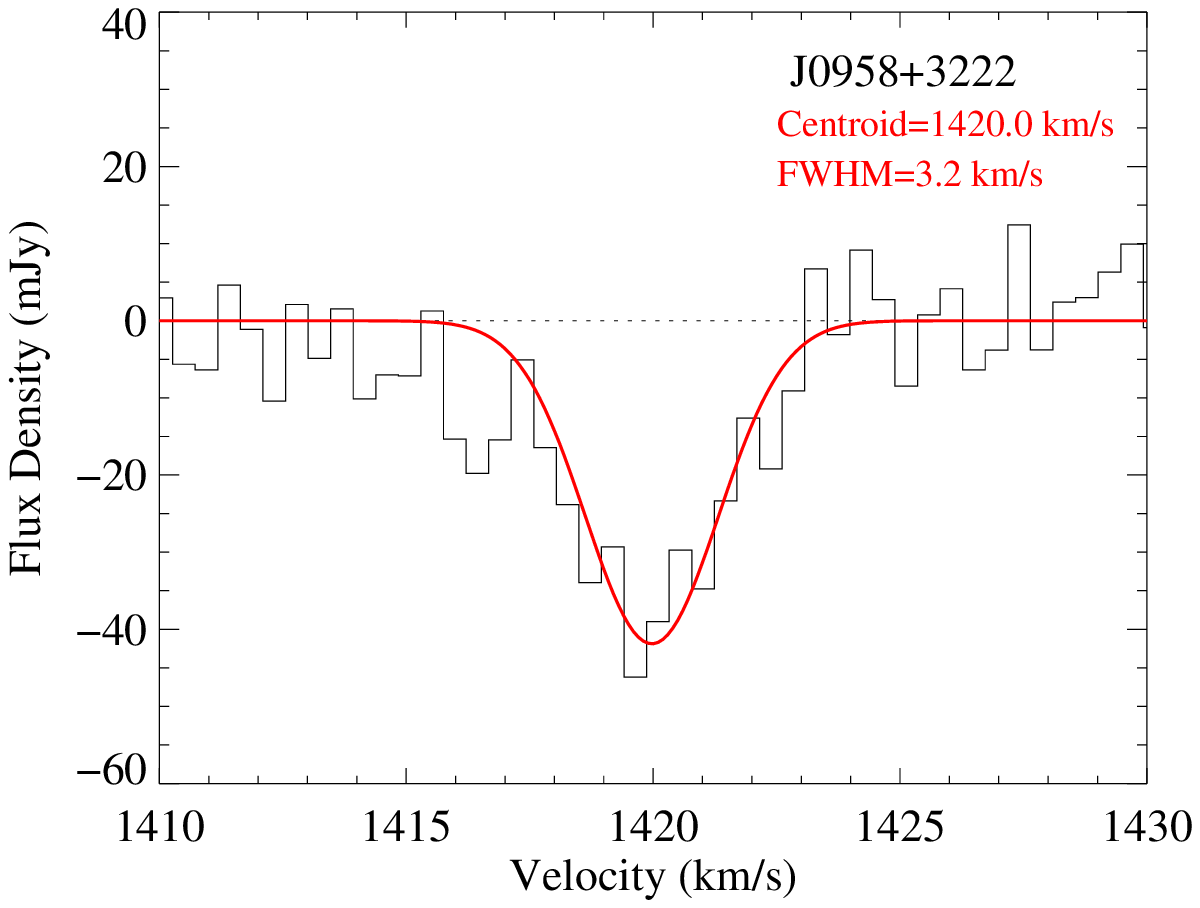}   
\caption{21cm \textsc{Hi}  absorption observed towards sightline J0958+3224 probing galaxy J0958+3222 at $z=0.0049$. The left panel shows the \textsc{Hi} profile with broad emission associated with the host galaxy with the absorption towards the quasar sightline at 1420~\kms over laid on the emission profile. The data is binned to 2~\kms. The right panel is the  emission subtracted absorption profile. The red curve is the fit to the data corresponding to a FWHM of 3.2 \kms. The spectra is binned to a spectral resolution of 0.5~\kms.}
 \label{fig-hi_abs} 
\end{figure}

\section{ Results and Implications \label{sec:Gass_halos}}

\subsection{Data: Overview}

We obtained science quality data for 15 of the 16 sightlines from our sample. Data for sightline towards J1228+3706 probing galaxy J1228+3706 at redshift of 0.1383 were severely corrupted by RFI. 
Therefore, we exclude this sightline from our analysis in this section. 
We detected one 21~cm \HI absorption feature among those 15 sightlines where we had good data. Thus, our 21~cm \HI absorption detection rate is about 7\%. This is substantially lower than those previously published by G10, B11, \citet{srianand13}. However, our detection rate is consistent with the findings of \citet[][R15 hereafter]{reeves15}, who reported a detection rate of 4\%. We discuss the possible reasons behind the discrepancy in the detection rates in the next section.

We detected \HI in absorption towards sightline J0958$+$3224 probing foreground galaxy J0958+3222, also known as NGC~3067. This absorber was first discovered by \citet{haschick75} and subsequently re-observed by \citet{carilli92, keeney05}. Figure~\ref{fig-hi_abs} shows the absorption feature over the \HI emission profile from the galaxy in the left panel and an emission subtracted profile in the right panel. The data were binned to a resolution of  2~\kms and 0.5~\kms in the left and the right panels respectively.
The full width at half maximum (FWHM) of the absorption feature shown in the right panel is 3.2($\pm$0.5)~\kms and a peak depth of the feature is 42~mJy corresponding to a peak optical depth of  0.035. The peak depth drops to 41~mJy when the data is smoothed to 1~\kms. The optical depth(s) is consistent with that found by \citet{keeney05} near the core of the background source in the high spatial resolution data obtained with the  Very Large Baseline Array (VLBA) with a spectral resolution of 1.7~\kms. It is worth pointing that our data represents the highest resolution spectra of this system and we would have detected sub-\kms sub-structures if there were any (at optical depths $>$  0.01 when binned to 1~\kms.)

In most cold clouds the \HI 21~cm hyperfine transition is the result of collisional excitation. Therefore, the temperature of the cloud is directly related to the width of the line. Baring no contribution from turbulence, the spin temperature, $T_s$, should be the same as the kinetic temperature, $T_k$, which in turn can be related to the full-width at half maximum (FWHM) of the absorption feature as
 \begin{equation}{\label{eq-kin_temp}}
 T_{k}\le 21.855~(\Delta ~v)^{2}
 \end{equation}
where $\Delta v$ is the FWHM velocity in \kms. Based on this relationship, we estimate the spin temperature, $T_s$ of the cold \HI towards J0958$+$3224  to be 224$^{+73}_{-64}$~K. This should be considered the upper limit as the width of the line may have some contribution from turbulence, which would mean a lower spin temperature. Therefore, $\rm T_s \le 224~K$.

The column density of the \HI\ can be estimated using the following equation \citep{rohlfs86}:
\begin{equation}{\label{eq-col_den}}
N({\rm H~I}) = 1.823 \times 10^{18} \frac{T_{s}}{f}\int \tau(v) dv,
\end{equation}
where $\tau(v)$ is the 21~cm optical depth of as a function of velocity in km~s$^{-1}$ and $f$ is the covering fraction  of the background source by the absorbing cloud.  Assuming $f=1$ and $T_s \le$ 224~K, the column density of the absorber is $\le \rm 4.4 (\pm 1.5)\times 10^{19}~cm^{-2}$. The error cited is the uncertainly in the estimation of the column density, however, since $\rm T_s \le T_k$ this estimate is a lower limit. Our spin temperature estimate is about 50\% lower than that estimated by \citet{keeney05} based on the comparison of 21cm \HI feature and \Lya absorption associated with this system. Since \Lya absorption traces the total neutral hydrogen and 21~cm \HI absorption is most sensitive to the cold ($\lesssim 300$~K) component of the neutral hydrogen, the difference in the estimated spin temperature suggests that the cold gas to total neutral gas is 44($\pm 18)$\%. Again, the estimate of the cold gas fraction assume that $\rm T_s = T_k$. Therefore, if turbulence in the cloud has significant contribution to the line-width, then the cold gas fraction would be much smaller.

In the other 14 sightlines from our sample, we did not detect any absorption feature. The limiting optical depths corresponding to 3$\sigma$ noise are presented in Table~2. These were estimated by binning the raw spectra to 1~\kms from its original resolution, which is an order-of-magnitude higher. We also created a large suit of binned data at resolutions of 1, 3, and 5~\kms to look for weak and broad absorption features. However, we found none. Therefore, despite galaxies in our sample being probed at small impact parameters, we conclude that detection rate of 21~cm \HI absorption is $\sim$7\% for the entire sample. We discuss the implication of this finding and possible caveats in the next section.

\begin{figure*}
\includegraphics[trim = 0mm 0mm 0mm 0mm, clip,scale=0.65,angle=-0]{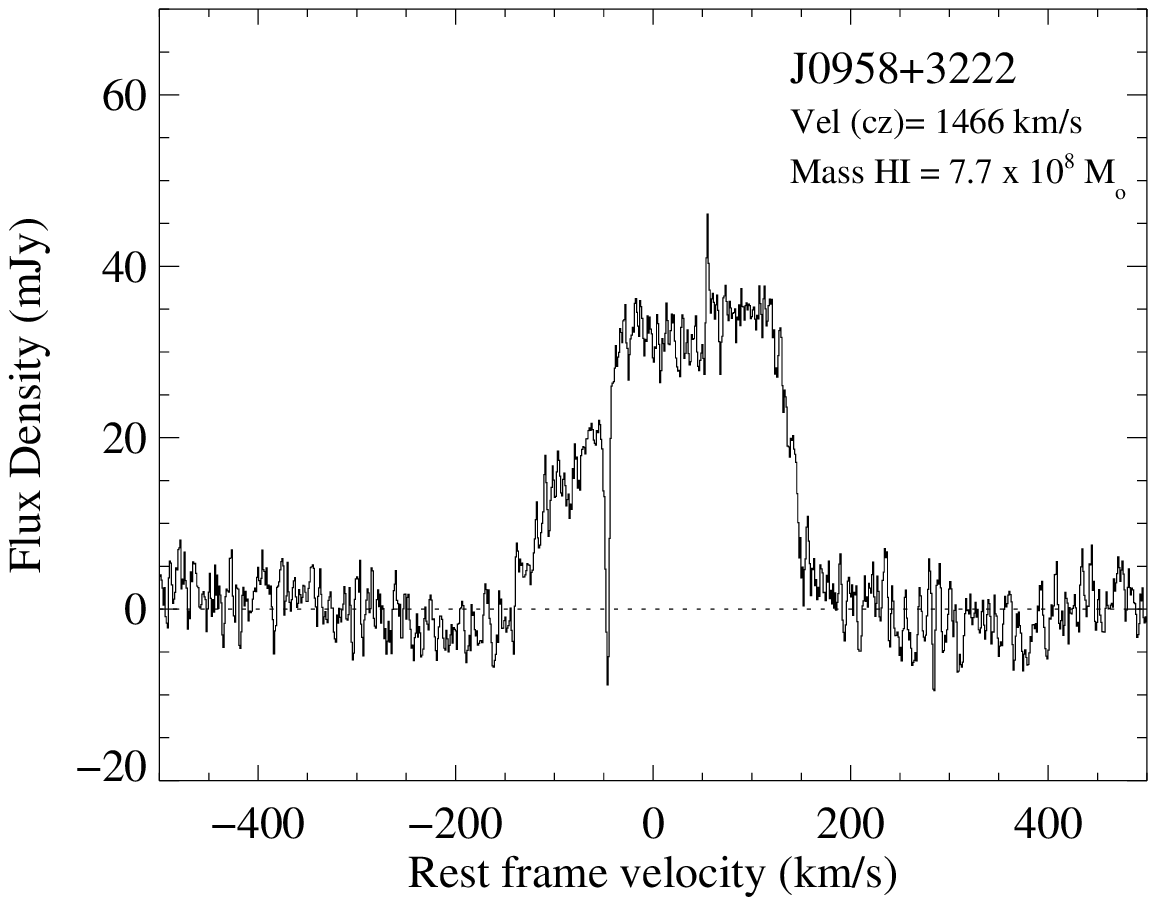}   
\includegraphics[trim = 0mm 0mm 0mm 0mm, clip,scale=0.65,angle=-0]{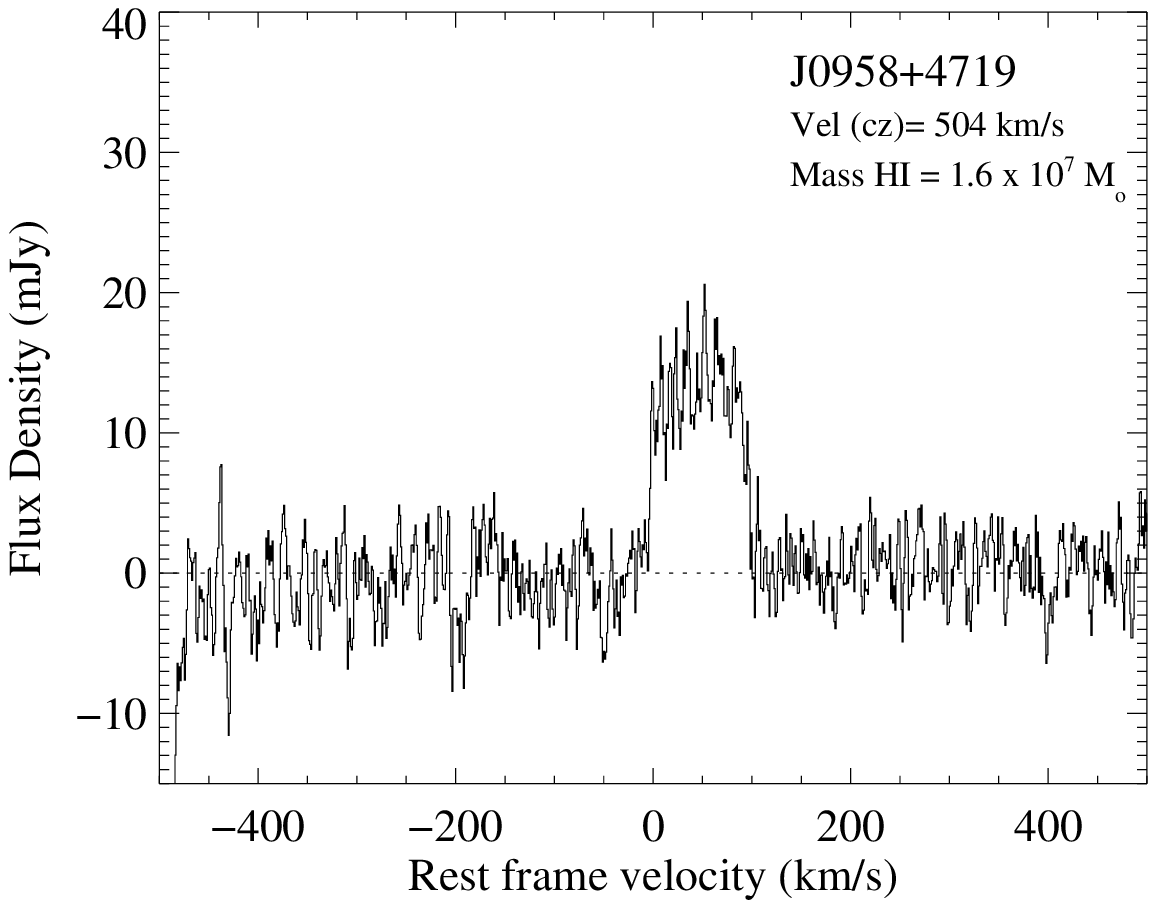}   \\ 
\includegraphics[trim = 0mm 0mm 0mm 0mm, clip,scale=0.65,angle=-0]{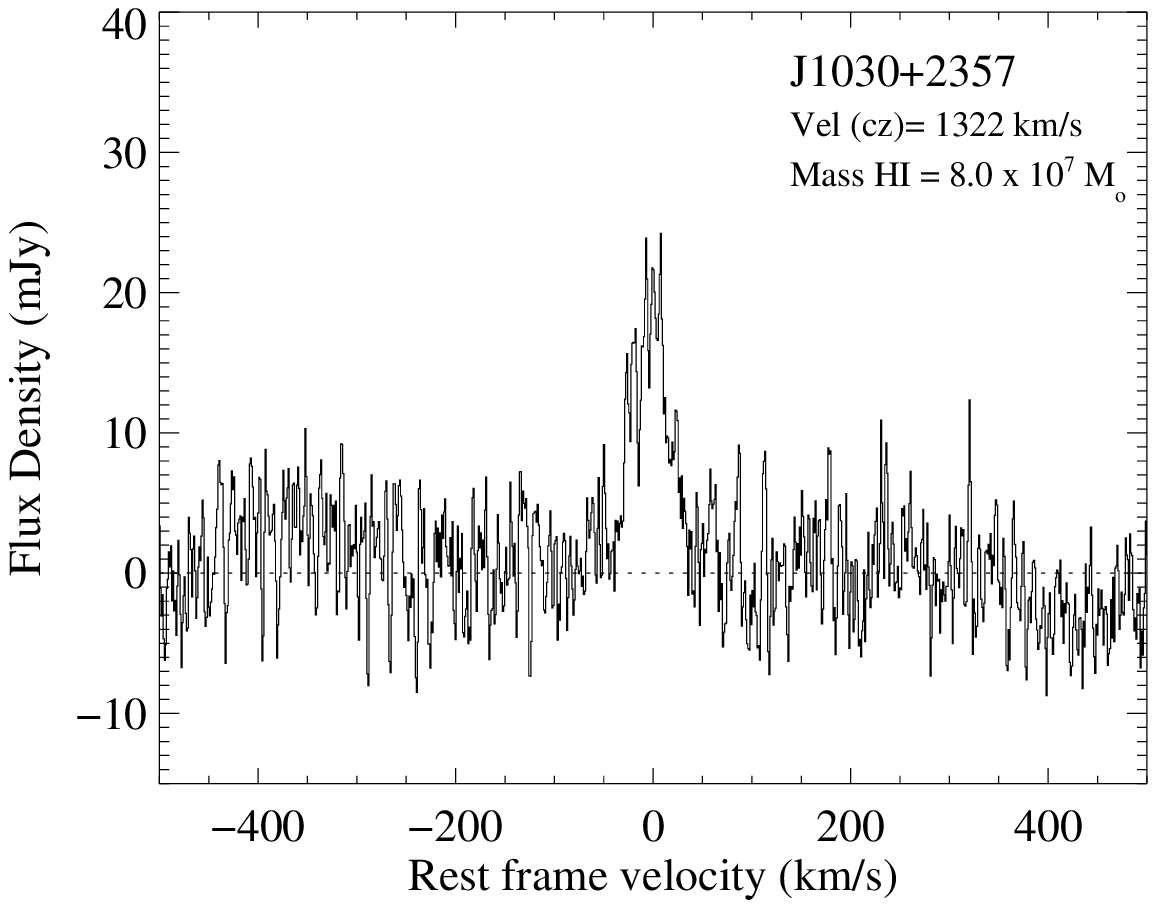}   
\includegraphics[trim = 0mm 0mm 0mm 0mm, clip,scale=0.65,angle=-0]{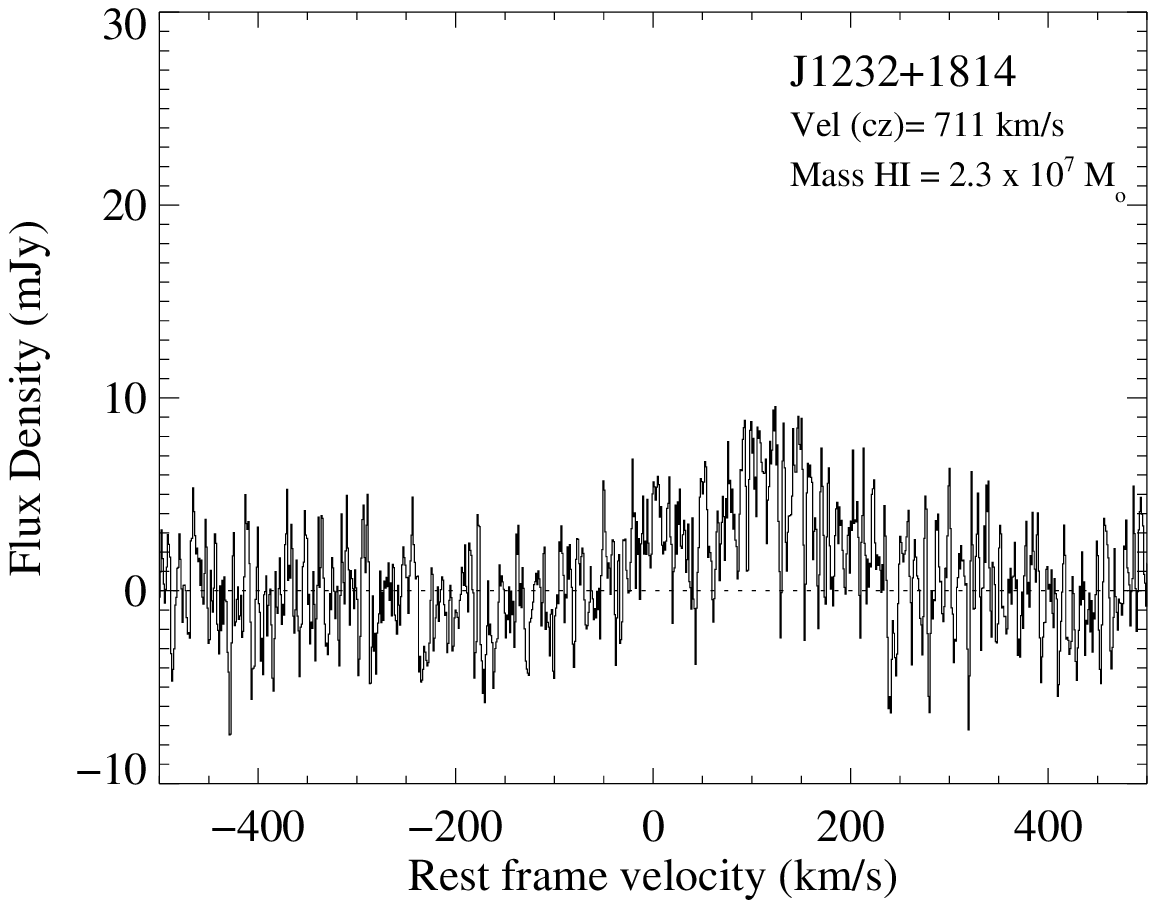}  \\
\includegraphics[trim = 0mm 0mm 0mm 0mm, clip,scale=0.65,angle=-0]{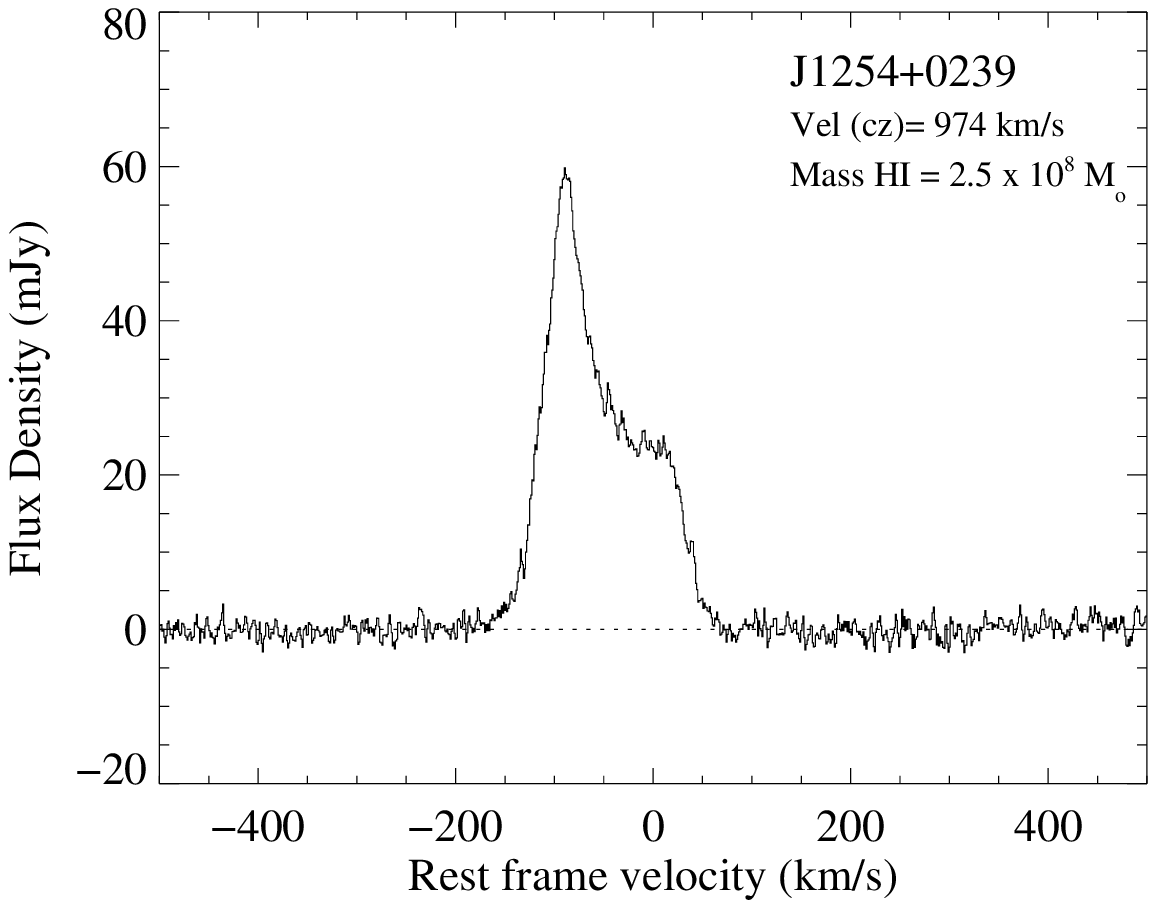}   
\includegraphics[trim = 0mm 0mm 0mm 0mm, clip,scale=0.65,angle=-0]{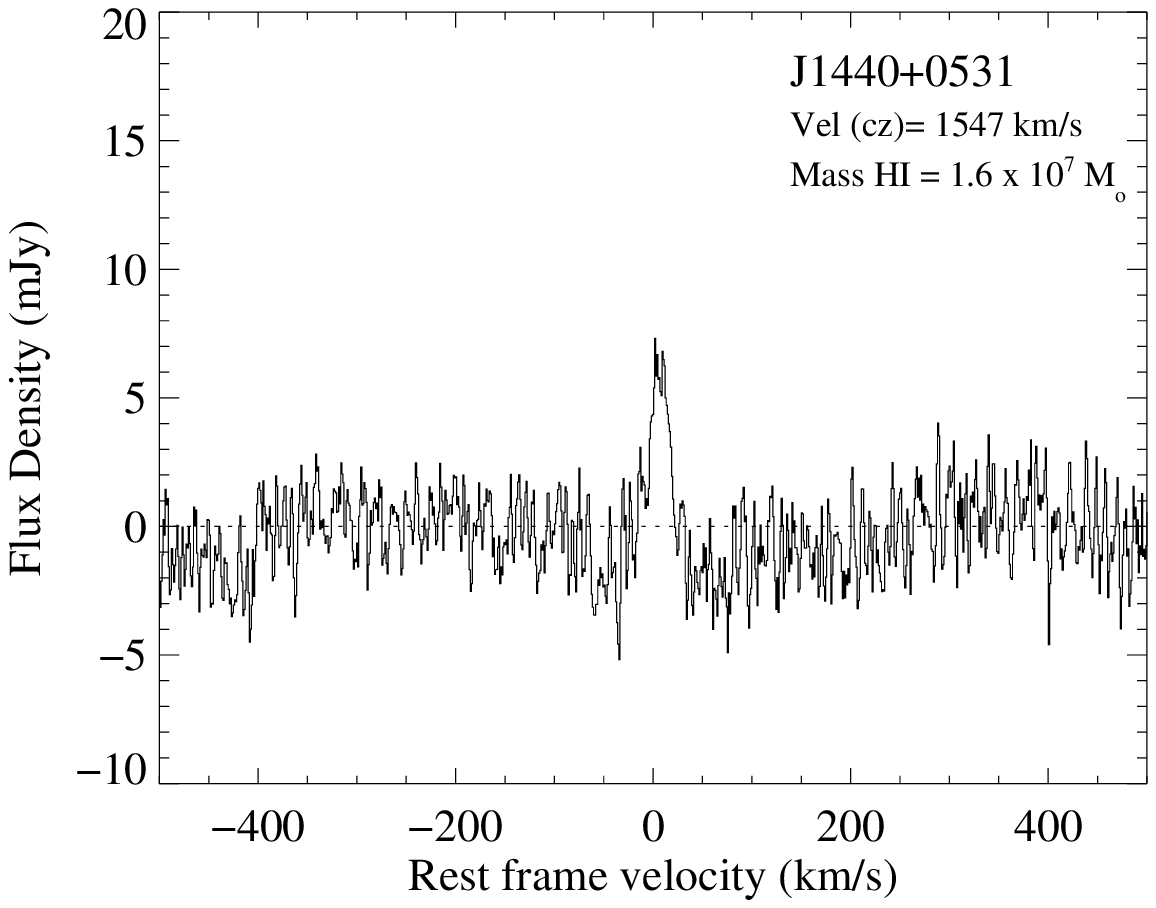}  
\caption{Spectra of the 6 galaxies where \textsc{Hi} was detected in emission. The spectra are shown in the rest-frame velocity units with zero corresponding to the SDSS optical redshift of the stellar body of the galaxy. The \textsc{Hi} masses associated with the galaxies are printed in the top right corner of the plots. The sensitivity of the data varies between the different spectra, which depend on the luminosity distance and the observing time.  }
 \label{fig-HI_emission}
\end{figure*}

We detected \HI emission in 6 of the 15 galaxies. The \HI spectra are presented in Figure~\ref{fig-HI_emission}. Their \HI masses range from $\rm 1.6\times 10^7 - 7.7\times10^8~M_{\odot}$. Since the survey was not designed to detected \HI in emission, the detectability limit for \HI in emission varies widely from galaxy to galaxy. Therefore, the limits on the \HI masses for the remaining galaxies varies widely from $0.01-23.72\times 10^8~M_{\odot}$. This is a result of varying S/N in the spectra as well as the large range in luminosity distances (corresponding to redshift  0.0013$<z<$0.1383). Based on the \HI mass and the limits, we estimate the size of the \HI disks, $\rm R_{HI}$, in these galaxies following the prescription by \citet{swaters02}. The sizes (radius of the disks) are presented in Table~2. Only one of the sightlines is expected to probe the \HI disk of the foreground galaxy. In most of the other, cases we are probing the regions associated with the extended \HI disk and extraplanar gas of the target galaxies.

\begin{figure*}
\includegraphics[trim = 0mm 0mm 40mm 0mm, clip,scale=0.60,angle=-0]{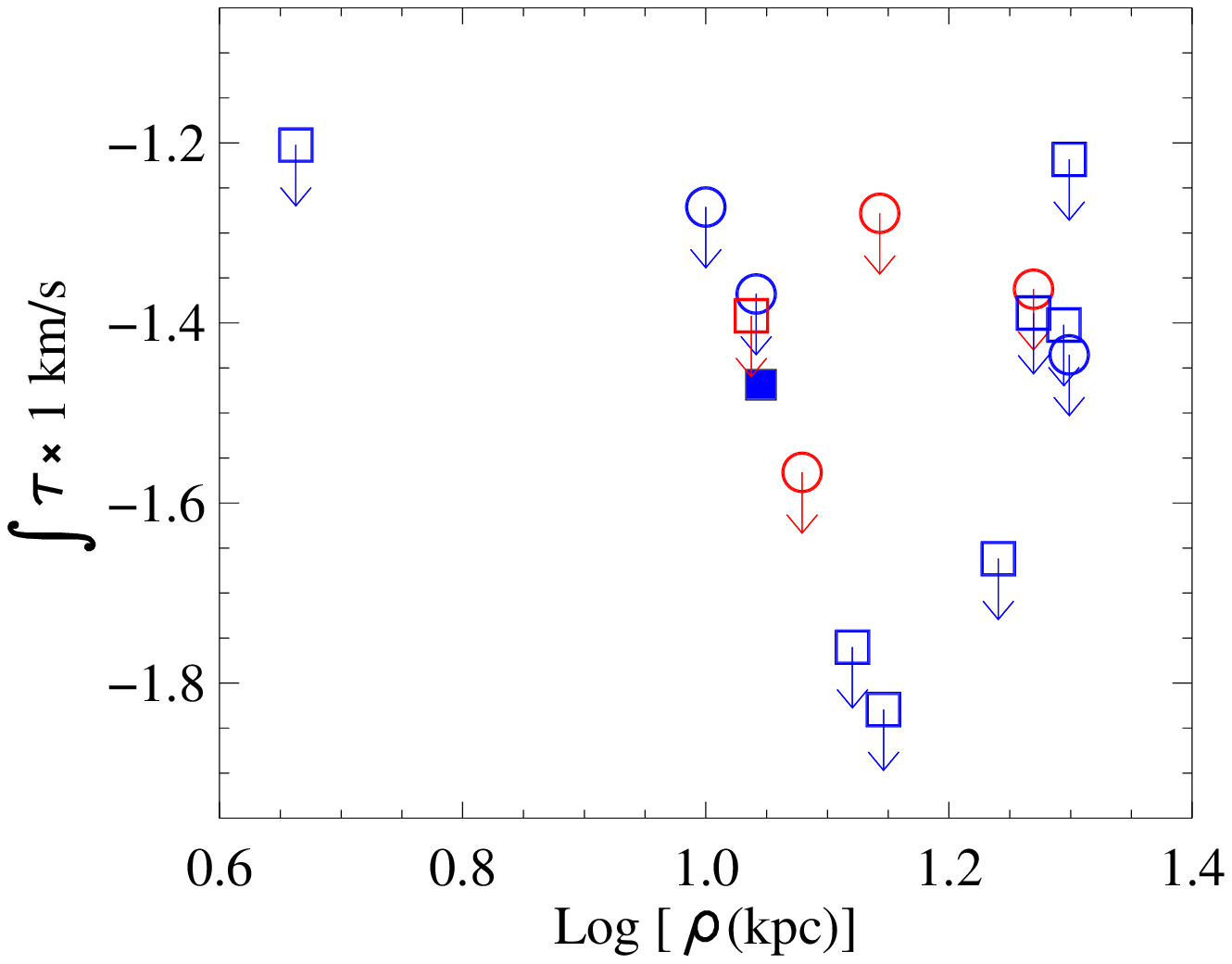}  
\includegraphics[trim = 0mm 0mm 00mm 0mm, clip,scale=0.60,angle=-0]{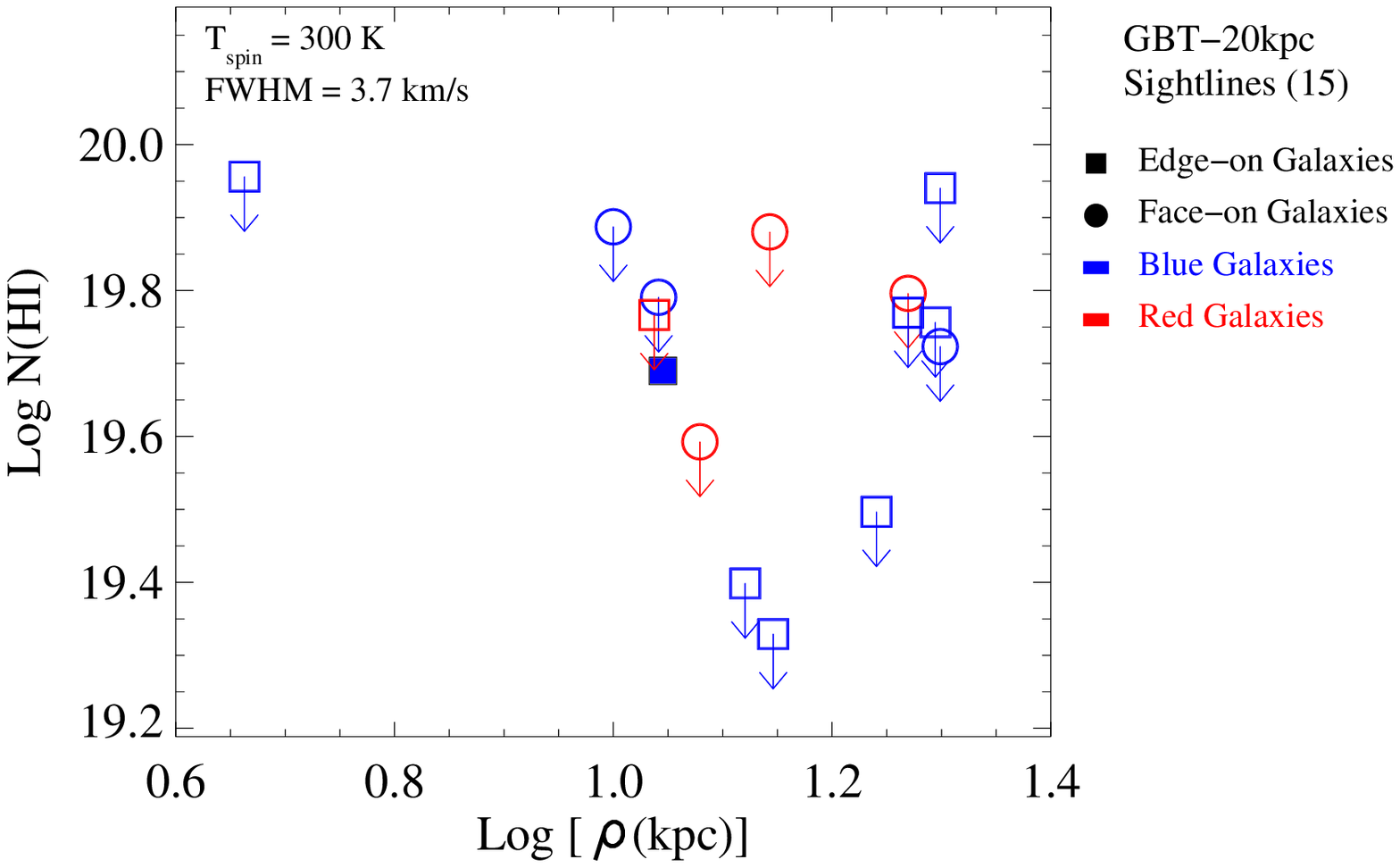}  
\caption{Left: The measured optical depths for our sample are plotted  as a function of impact parameter. The filled symbol represents the detection. The color of the symbols represents the color of the galaxy as discussed in Section 3.1. The orientations of the sightlines are indicated via the shape of the symbols.  The limiting values refer to 3$\sigma$ optical depth that was estimated by integrating the noise over 1~\kms or $\sim$12-14 channels. 
Data for sightlines probing galaxy J1228+3706 were severely corrupted by RFI and hence no measurements could be made for this sightline. Right: Column densities are plotted as a function of impact parameter assuming a spin temperature of 300~K corresponding to a FWHM of $\sim$3.7~\kms and a covering fraction of 100\%. For lower spin temperatures the column densities will be proportionally lower.}
 \label{fig-tau_rho} 
\end{figure*}

We present the optical depths observed for the sample as a function of impact parameter, $\rho$, in Figure~\ref{fig-tau_rho}. The detection towards J0958$+$3224 is shown as the filled square. The color and the symbols indicate the color and orientation (face-on vs. edge on) of the host galaxies. The galaxy colors were defined using the empirical result from \citet[][equation A6]{vandenbosh08} that uses SDSS {\it g-r} band k-corrected magnitudes to 
split the galaxy population into red passive galaxies and blue star-forming galaxies (see B11 for more details). For the smallest  impact parameter sightline probing J1440+0531 at 4.6~kpc, we estimate a 3$\sigma$ limiting peak optical depth of 0.063. This is close to the limiting values we chose while designing the program. However, for quarter of the sightlines our limits are less than half of this value. We also plot the \HI column density, N(HI), as a function of impact parameter in the right panel of Figure~\ref{fig-tau_rho}. The column densities in the case of non-detections were estimated assuming a Gaussian profile for the absorption feature with peak optical depth corresponding to the 3$\sigma$ limit on the observed optical depth and a spin temperature of 300~K. The column densities range in the sub-damped \Lya absorber (sub-DLAs ; defined by \citet{peroux01} as  1.6$~ \times ~10^{17}$ cm$^{-2}~ <~  N$(\ion{H}{1}) $< 2 \times 10^{20}$ cm$^{-2}$). Again, we do not see any correlation of column density with impact parameter. This is not surprising as the column densities were derived directly from the limiting optical depths.

 \begin{figure*}
\includegraphics[trim = 0mm 0mm 40mm 0mm, clip,scale=0.60,angle=-0]{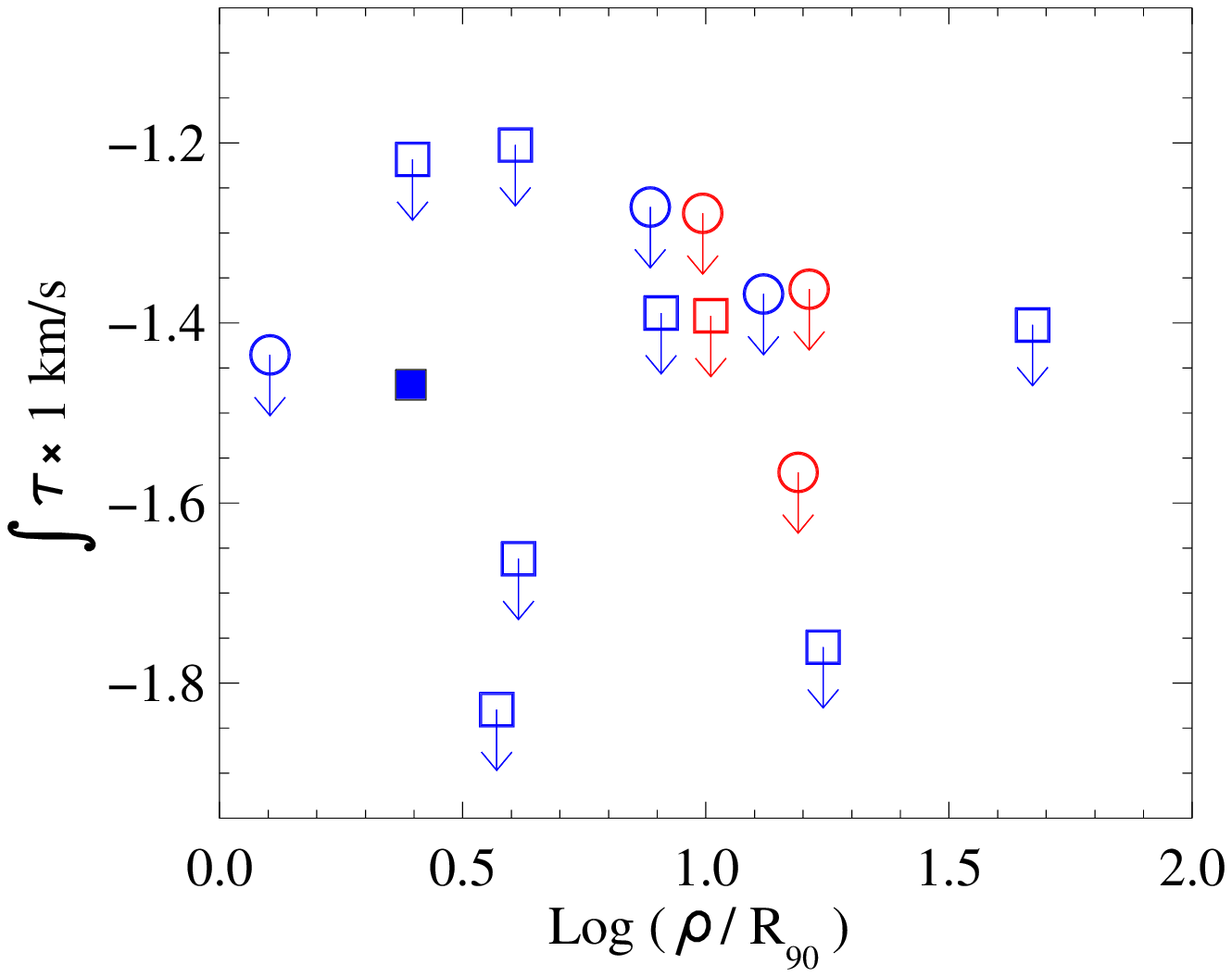} 
\includegraphics[trim = 0mm 0mm 00mm 0mm, clip,scale=0.60,angle=-0]{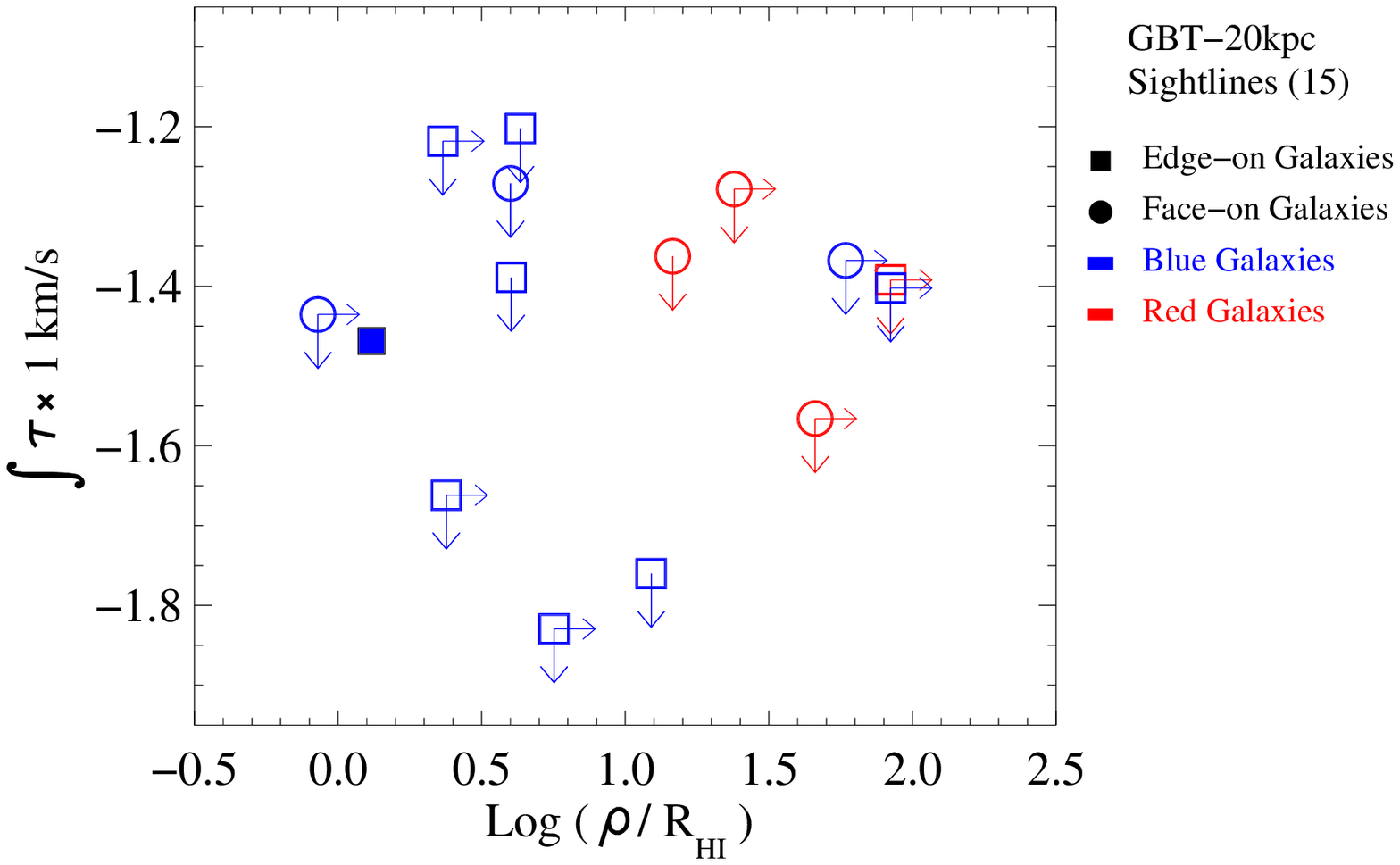} 
\caption{The measured optical depth and limiting optical depths are plotted as a function of normalized impact parameter. The filled symbol represents our detection. The color of the symbols represents the color of the galaxy as discussed in Section 3.1. The orientations of the sightlines are indicated via the shape of the symbols. The limiting values refer to 3$\sigma$ optical depth that was estimated by integrating the noise over 1~\kms or $\sim$12-14 channels. Data for sightlines probing galaxy J1228+3706 were severely corrupted by RFI and hence no measurements could be made for this sightline. The left panel shows the optical depth as a function of impact parameter normalized by the optical size of the disk as defined by the SDSS r-band petrosian radius (R90). The right panel shows optical depth as a function of impact parameter normalized by the of \textsc{Hi} radius, $\rho/ \rm R_{HI}$. \textsc{Hi} disk radius was estimate using the tight \textsc{Hi} mass to size relationship observed for galaxies. Here we use the prescription by \citet{swaters02} for the conversion.}
 \label{fig-tau_rho_norm} 
\end{figure*}

 \begin{figure}
\includegraphics[trim = 0mm 0mm 0mm 0mm, clip,scale=0.65,angle=-0]{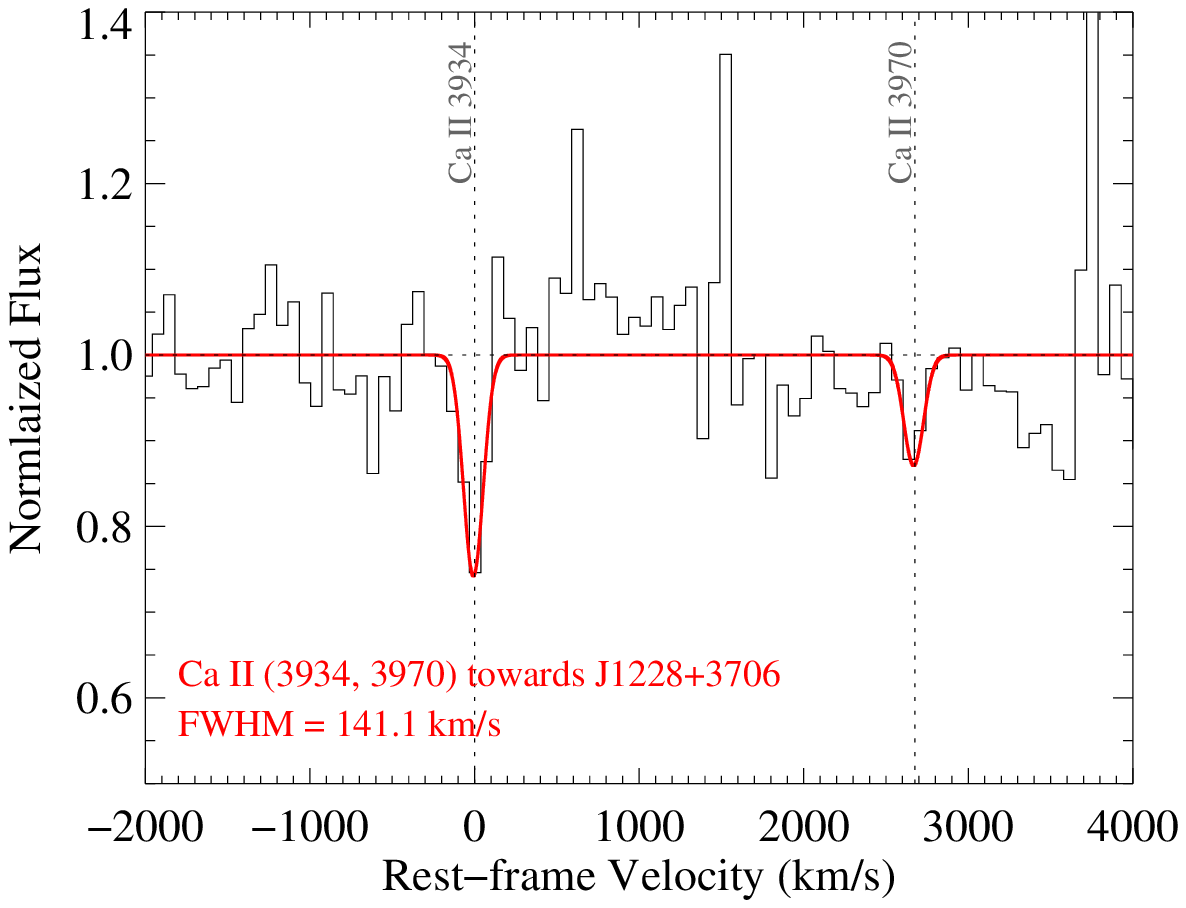} 
\caption{SDSS spectrum of optical QSO J1228$+$3706 showing \ion{Ca}{2} $\lambda\lambda$ 3933, 3969 $\rm \AA$ absorption at the redshift of the foreground galaxy J1228$+$3714 at $z=0.1383$. The red curve represents the best fit to the data. The lines in the doublet are optically thin and exhibit a ratio of 2:1.}
\label{fig-CaII} 
\end{figure}

This finding is inconsistent with the recent result by \citet{curran16} where they observed an inverse correlation of 21~cm \HI strength and impact parameter. However, it is worth mentioning that there sample consisted of a ``mixed" sample from various different surveys and some of the sightlines have been pre-selected due to existence of  metal-line absorbers of DLAs. Another reason for our sample not showing any correlation could be the small range of impact parameters ($\rho<20~\rm kpc$) probed in by our sample unlike the sample studied by \citet{curran16}.
The \HI absorber detected in this sample lies at an impact parameter of 11.1~kpc, which is close to the average of the impact parameter range. However, it is worth noting that the sample is fairly heterogeneous in terms of galaxy properties. We have dwarf galaxies with stellar masses of 10$^7~\rm M_{\odot}$ as well as L$_{\star}$ galaxies of the order of 10$^{10}~\rm M_{\odot}$. So the impact parameter may not be the right quantity to express the scale for these galaxies spanning more than three orders-of-magnitude in stellar mass. 

In order to find a suitable parameter to quantify the distance of the region probed by the sightline with respect to the disks of the galaxies, we defined normalized quantities. We chose the optical size (petrosian radius derived from the SDSS r-band image), $\rm R90$, and the \HI disk size, $\rm R_{HI}$ as the normalizing parameters. The variation of optical depth as a function of the normalized impact parameter in terms of the optical size $\rho/ \rm R_{90}$ and \HI size $\rho/ \rm R_{HI}$ are presented in Figure~\ref{fig-tau_rho_norm}. For galaxies where we did not detect \HI in emission, we quote an upper limit for $\rm R_{HI}$ and a lower limit for $\rho/ \rm R_{HI}$. These are shown a right pointing arrow in the right panel of Figure~\ref{fig-tau_rho_norm}. Again, we do not find any correlation between the detection rates or strengths of 21~cm \HI absorption and normalized impact parameter. Interestingly, the only detection in our sample does show a migration towards lower normalized parameter - both in terms of optical and \HI sizes. Although we have another sightline with lower/similar normalized impact parameter that do not show any absorption. This may indicate that there might be a weak correlation between cold \HI detected in \HI 21~cm absorption and distance from its host galaxy, but with a large dispersion. We also do not find any dependence of optical depth with the orientation of the sightline with respect to the foreground galaxy.

\subsection{Associated Metal-lines}

While our sample was not selected based on the optical properties of the background sources, some of them (12/16) do have SDSS spectra. Taking advantage of the existing spectra, we examined the range of wavelengths where associated \ion{Ca}{2} absorption would be present, if any. We detected \ion{Ca}{2} absorption features in  sightlines towards J0958$+$3224 and J1228$+$3706. The normalized spectra showing the \ion{Ca}{2} doublet detected towards the sightline J1228$+$3706 associated with foreground galaxy J1228+3714 at z = 0.1383 is presented in Figure~\ref{fig-CaII}. The best-fits to the data are plotted in red. The lines are optically thin and exhibits a 2:1 ratio as expected for the doublet in the optically thin regime. We note that the low spectral resolution of the SDSS spectrum has resulted in spearing of the lines. Unfortunately the \HI spectrum of this galaxy was severely corrupted by RFI and hence we could not make any measurements for 21~cm \HI absorption.

For details on the \ion{Ca}{2} absorber towards J0958$+$3224, we refer the reader to published work by \citet{haschick75, carilli92, keeney05}. 
  
 \section{Discussion}
 
 Based on our previous understanding from studies by G10, and B11, we expected to detect 21cm absorbers in 50\% of the galaxies in our sample. However, our detection rate is considerably smaller than that predicted by these studies. However, our detection rate is consistent with that measured by R15. The reason for the discrepancy among the conclusions between  our survey and that of G10 and B11 surveys is not absolutely clear. One possible reason could be the unbiased nature of the sample.
While the observational setup for our survey is almost identical to the survey by B11;  
unlike the B11 sample, our sample was not optically selected and the background radio sources included both optically bright and faint Quasi-stellar objects (QSOs).

Another possible cause for the low detection rate could be the large beam of the GBT (9.1$^{\prime}$ at 20~cm). This can lead to \HI absorption being filled-in my emission as in the case of UGC~7408 whose GBT spectrum did not show any absorption but its VLA D-configuration spectrum did show a narrow absorption feature (see B11 for details). This would be of concern in cases where we detect \HI in emission - that is in the 6/15 galaxies. For the remaining 9 sightlines, the beam filling would be of minimal concern. Only if all of these 6 sightlines show \HI absorption in high spatial resolution data (that we are planning to acquire) the detection rate would come to 40\%, which would be close to that seen by G10 and B11. One the other hand, our detection rate is consistent with that of R15, which would suggest that the covering fraction of cold \HI in the inner halo ($\rho<20$~kpc) of galaxies is very low.

Comparing this result to our own galaxy, we find our estimate of covering fraction of \HI to be broadly consistent with that seen in the halo of the Milky Way. \citet[][W91 hereafter]{wakker91} reported an \HI covering fraction of 18\% for HVCs with column density log~N(HI)$> 18.5$. Although this value of covering fraction for the Milky Way is higher  that what we measure in our survey (18\% vs. 7\%), the column density limit for the study by W91 is almost an order of magnitude lower. Hence, we expect that the value will drop if only higher column density regions are considered. On the other hand, W91 used \Lya  absorption to trace neutral hydrogen which is sensitive to both cold ($\lesssim$ 300~K) and warm ($\sim$1000~K) \HI. This is not true for 21~cm \HI absorption, which is most sensitive to the cold phase (T$\lesssim$300~K) of \HI. The warm \HI would produce a broad and shallow 21~cm feature that requires data with extremely high sensitivity and baseline stability to be able to detect it. So far none of the surveys have been able to achieve both of these criterion and hence were not able to probe to the warm component of \HI. Consequently, we expect the covering fraction from 21~cm studies probing cold \HI at column densities, log~N(HI) $\gtrsim 20.0$, to be that of less than that of ``total" \HI at column densities log~N(HI)$> 18.5$.

Therefore, the process of condensation of inflowing gas into cold atomic gas in galaxies does seem to happen very close to the galaxy (at radii $<<$ 20~kpc), although the condensation to warmer \HI ($\sim 1000$~K) might happen at larger distances. We also do not see any difference in the covering fraction between red and blue galaxies in the sample. However, many of the blue galaxies show \HI in emission in their spectra and hence might be hiding 21~cm \HI absorption features.

\section{Summary \& Implications \label{sec:conclusion}} 

One of the main phases of the baryon cycle in galaxies is the condensation of inflowing gas into cold ($\lesssim 300$~K) clouds that will eventually collapse to form stars. In order to investigate the process of condensation, we conducted the Green Bank Telescope survey to probe 21~cm \HI in absorption in the inner halos of low-$z$ galaxies. The sample was selected by cross-correlating the Quasars from the VLA FIRST Survey and galaxies from the Sloan Digital Sky Survey (SDSS) to find galaxy-quasar pairs. The quasar sightlines were chosen to probe the foreground galaxy at an impact parameter of $\le$ 20~kpc. Furthermore, in an attempt to optimize the observing time, we selected sightlines with 20~cm radio fluxes of background sources  greater than 100~mJy. This yielded 16 galaxy-quasar pairs probing  galaxies with luminosities from 0.0001 to 1.5~L$_{*}$ in the redshift range 0.001 $\le$ z $\le$ 0.18. 

We observed the sample with the Green bank Telescope in the L-band (20~cm) that resulted in obtaining good data for 15 of the 16 sightlines. We detected one \HI absorber towards sightline J0958$+$3224 associated with foreground galaxy  J0958+3222, also known as NGC~3067. 
We also detected \HI in emission from six of the galaxies. We also analyzed the archival SDSS data for 12/16 of the background quasars, where optical spectra were available. We detected two \ion{Ca}{2} absorbers associated with galaxies J0958+3222 and J1228+3706. 

Based on our analysis we came to the following conclusions:
\begin{itemize}
\item[1.] The covering fraction of cold \HI ($\lesssim 300$~K) is $\sim$7\% for neutral gas of column density $\gtrsim \rm 10^{20}~cm^{-2}$ in the inner halo ($\rho<$20~kpc) of galaxies. Our estimate is substantially different from that previously observed by G10 and B11 of 50\%. However, it is consistent with the covering fraction of 4\% that was reported by R15. This low covering fraction is also broadly consistent with the covering fraction of 18\% for HVCs at column densities of log~N(HI)$> 18.5$.
\item[2.] The most likely cause of the discrepancy in the covering fraction between B11 and our result is the nature of the sample. The observational setup and data reduction procedure was almost identical. While our sample was radio selected, the B11 sample (and so is the G10 sample) was selected based on the optical and radio properties of the background quasars.
\item[3.] We detected an \HI absorber towards the sightline J0958$+$3224 probing foreground galaxy J0958+3222. This absorber was observed by multiple studies including \citet{keeney05}. Our observations are the highest spectral resolution observations of this system. The measured FWHM of the absorber was found to be 3.2($\pm$0.5)~\kms with a peak optical depth of 0.035. The optical depth is consistent with that found by \citet{keeney05}. Based on the FWHM, we estimate the spin temperature to be  $\le$ 224~K and a column density of $\le \rm 4.4 \times 10^{19}~cm^{-2}$. 
A comparison of our spin temperature estimate for the cold component of the \HI in this system to that of the total \HI as estimated by \citet{keeney05} suggests that the cold gas to total neutral gas is about 44($\pm 18)$\%.
\item[4.] We do not find any correlation in optical depth or column density of 21~cm \HI absorption with impact parameter for absorbers with optical depths of $\geqslant 0.06$. Furthermore, no correlation was observed between optical depths and  normalized impact parameters $\rho$/R$_{90}$ and $\rho$/R$_{HI}$.
\item[5.] We also do not find any dependence of 21~cm \HI optical depth with the orientation of the sightline with respect to the foreground galaxy.
\item[6.] We also searched for \ion{Ca}{2} in the archival SDSS spectra that were available for 12/16 background quasars. \ion{Ca}{2} absorption features were detected at the redshift of the target galaxies J0958+3222 and  J1228$+$3706. The associated \HI absorber for  J0958+3222 was detected in 21~cm, no such measurements could be made for J1228$+$3706 as our spectral-line data for this sightline were severely corrupted by RFI.

\end{itemize}

Our characterization of the distribution of \HI in the inner halos of galaxies indicates that the process of condensation of inflowing gas into cold \HI does occur at the very inner regions of the galaxies. However, evidence from \HI emission maps of the Milky Way and several other nearby galaxies suggest that the neutral gas in the halos might be existing mostly in the warm phase at temperatures of of-the-order of 1000~K. Further observations are required to precisely estimate the covering fraction of \HI - especially for the galaxies exhibiting \HI emission. We plan to carry out higher spatial resolution observations in the near future with the VLA. For future studies, we recommend sensitive observations with high spectral resolution of sub-\kms and high spatial resolution optimized to the size of the background source for maximizing the success of detecting cold gas in the inner halos of galaxies.

\vspace{.5cm}
\acknowledgements 
I thank the referee for his/her constructive comments. 
My sincere thanks to the staff at the Green Bank Telescope for the support during the observations and data reduction. 
I thank Neeraj Gupta, Nissim Kanekar, R. Srianand, and Elaine Sadler for general discussions on the topic of this paper.
I would also like to thank my collaborators from our previous 21~cm \HI study: Min Yun, Emmanuel Momjian, Timothy Heckman, Todd Tripp, David Bowen, and Donald York for useful discussions during that program that led to the motivation for this study.
Finally, I extend my thanks to Leo Blitz for hosting me at UC Berkley.

This project also made use of SDSS data. Funding for the SDSS and SDSS-II has been provided by the Alfred P. Sloan Foundation, the Participating Institutions, the National Science Foundation, the U.S. Department of Energy, the National Aeronautics and Space Administration, the Japanese Monbukagakusho, the Max Planck Society, and the Higher Education Funding Council for England.  The SDSS Web Site is http://www.sdss.org/. 
The SDSS is managed by the Astrophysical Research Consortium for the Participating Institutions. The Participating Institutions are the American Museum of Natural History, Astrophysical Institute Potsdam, University of Basel, University of Cambridge, Case Western Reserve University, University of Chicago, Drexel University, Fermilab, the Institute for Advanced Study, the Japan Participation Group, Johns Hopkins University, the Joint Institute for Nuclear Astrophysics, the Kavli Institute for Particle Astrophysics and Cosmology, the Korean Scientist Group, the Chinese Academy of Sciences (LAMOST), Los Alamos National Laboratory, the Max-Planck-Institute for Astronomy (MPIA), the Max-Planck-Institute for Astrophysics (MPA), New Mexico State University, Ohio State University, University of Pittsburgh, University of Portsmouth, Princeton University, the United States Naval Observatory, and the University of Washington.

{\it Facilities:} \facility{GBT ()} \facility{Sloan ()}

\bibliographystyle{apj}	        
\bibliography{myref_bibtex}		

\clearpage
\begin{landscape}
\begin{deluxetable}{clrrrrlrrr cr ccc   ccccc}   
\tabletypesize{\scriptsize}
\tablecaption{Description of The Sample with Information on the Background Radio Sources and the Foreground Galaxies\label{tbl-gbt_sample}}
\tablewidth{0pt}
\tablehead{
\colhead{Sl \#} & \colhead{Radio Source} & \colhead{R.A.} & \colhead{Decl.} & \colhead{z$\rm _{radio}^{a}$} & \colhead{$S_{\rm 1.4GHz}$} & \colhead{Galaxy} & \colhead{R.A.} & \colhead{Decl.} & \colhead{z$\rm _{gal}^{b}$}   & \colhead{M$_{\star}\rm ^c$} & \colhead{Lumin$\rm ^c$} & \colhead{Color$\rm ^c$} &  \colhead{Impact Param.} &  \colhead{Orientation$\rm ^d$} \\
\colhead{} & \colhead{} &\colhead{} & \colhead{}  &\colhead{}  & \colhead{(mJy)} &\colhead{} &\colhead{} & \colhead{} & \colhead{}  &\colhead{(Log~[M$_{\odot}$])}  &\colhead{($\rm L_{\star}$)}&\colhead{}   &\colhead{(kpc)}&\colhead{}    }
\startdata
       1 & J0958+3224 & 149.587 & 32.401 & 0.54 &     1204 & J0958+3222 & 149.588 & 32.370 & 0.0049 & 10.1 & 0.5000 & Blue & 11.1 &           76  \\
       2 & J1203+0414 & 180.841 & 4.239 & 1.21 &     1123 & J1203+0413 & 180.845 & 4.230 & 0.0202 & 9.1 & 0.0311 & Blue & 14.0 &           77  \\
       3 & J0958+4725 & 149.582 & 47.419 & 1.85 &      763 & J0958+4719 & 149.504 & 47.324 & 0.0017 & 7.1 & 0.0009 & Blue & 13.2 &           87  \\
       4 & J1257+3229 & 194.489 & 32.492 & 2.29 &      589 & J1257+3228 & 194.485 & 32.482 & 0.0233 & 9.0 & 0.0370 & Blue & 17.4 &           47  \\
       5 & J1228+3706 & 187.198 & 37.103 & 1.48 &      382 & J1228+3714 & 187.117 & 37.235 & 0.0013 & 7.3 & 0.0013 & Blue & 11.0 & 0$\rm ^d$  \\
       6 & J1228+3706 & 187.198 & 37.103 & 1.48 &      382 & J1228+3706 & 187.199 & 37.102 & 0.1383 & 10.7 & 1.1669 & Blue & 15.2 & 0$\rm ^d$  \\
       7 & J0849+5108 & 132.492 & 51.141 & 2.72 &      344 & J0849+5108 & 132.490 & 51.145 & 0.0734 & 10.6 & 1.5141 & Blue & 19.9 & 0$\rm ^d$  \\
       8 & J1030+2359 & 157.697 & 23.994 & 0.34 &      336 & J1030+2357 & 157.707 & 23.965 & 0.0044 & 8.3 & 0.0093 & Blue & 10.0 & 0$\rm ^d$  \\
       9 & J1232+1809 & 188.058 & 18.157 & 0.09 &      307 & J1232+1814 & 188.080 & 18.245 & 0.0024 & 8.3 & 0.0064 & Red & 18.6 & 0$\rm ^d$  \\
      10 & J1221+3051 & 185.476 & 30.863 & 2.19 &      291 & J1222+3053 & 185.570 & 30.890 & 0.0018 & 7.2 & 0.0005 & Red & 12.0 & 0$\rm ^d$  \\
      11 & J1159+4412 & 179.825 & 44.205 & 1.16 &      236 & J1158+4411 & 179.737 & 44.193 & 0.0023 & 8.0 & 0.0038 & Red & 10.9 &           89  \\
      12 & J1218+1738 & 184.694 & 17.638 & 1.87 &      179 & J1218+1743 & 184.674 & 17.719 & 0.0023 & 8.3 & 0.0065 & Red & 13.9 & 0$\rm ^d$  \\
      13 & J1254+0233 & 193.689 & 2.558 & 0.32 &      152 & J1254+0239 & 193.713 & 2.654 & 0.0033 & 8.1 & 0.0081 & Blue & 18.6 &           53  \\
      14 & J1321+3510 & 200.482 & 35.177 & 1.58 &      144 & J1322+3512$^e$ & 200.615 & 35.205 & 0.0021 & 6.4 & 0.0002 & Blue & 19.7 &           78  \\
      15 & J1440+0531 & 220.133 & 5.523 & 0.81 &      114 & J1440+0531 & 220.115 & 5.532 & 0.0052 & 7.0 & 0.0012 & Blue & 4.6 &           86  \\
      16 & J1015+1637 & 153.811 & 16.628 & 0.93 &      110 & J1015+1637 & 153.809 & 16.633 & 0.0544 & 9.7 & 0.2063 & Blue & 19.9 &           39  
\enddata
\tablenotetext{a}{Photometric redshift from the SDSS. }
\tablenotetext{b}{Spectroscopic redshift from the SDSS. }
\tablenotetext{c}{Stellar mass, luminosity, and color defined based on NYUVAGC \citet{nyuvagc} and prescription by \citet{vandenbosh08}.  }
\tablenotetext{d}{Orientation is defined from SDSS position angle (see section 2.1 for details). Face-on galaxies have been assigned an angle of 0 degrees. }
\tablenotetext{e}{Pair of interacting galaxies. }
\end{deluxetable}
\clearpage
\end{landscape}

\clearpage

\begin{landscape}
\begin{deluxetable}{ccccccccccccccccc} 
\tabletypesize{\scriptsize}
\tablecaption{GBT 21~cm Observations and Measurements of \textsc{Hi}  Optical Depths and Masses For our Sample. \label{tbl-gbt_obs}}
\tablewidth{0pt}
\tablehead{
\colhead{Sl \#} & \colhead{Radio Source} & \colhead{$S_{\rm 1.4~GHz}$} & \colhead{Chan. Width} & \colhead{Gain} &  \colhead{Noise$^a$} & \colhead{$\tau$\tablenotemark{b}} & \colhead{Galaxy} &  \colhead{Impact Param.} &\colhead{Orient.}  & \colhead{Lum Dist}  &\colhead{HI Mass}   &\colhead{HI Radius}  &\colhead{R90$_r$} \\
\colhead{} & \colhead{} &  \colhead{(mJy)} & \colhead{($\rm km~s^{-1}$) } &\colhead{(K~Jy$^{-1}$)} & \colhead{(mJy)} & \colhead{} & \colhead{} & \colhead{(kpc)} & \colhead{} & \colhead{(Mpc)} & \colhead{($\times 10^8 M_{\odot}$)}  & \colhead{(kpc)} & \colhead{(kpc)} }
\startdata
       1 & J0958+3224 &     1204 & 0.076 & 1.85 & 4.7 & 0.034 & J0958+3222 & 11.1 &           76  &  21 & 7.72 $\pm$ 0.08& 8.5 & 4.5  \\
       2 & J1203+0414 &     1123 & 0.079 & 1.85 & 5.5 &  $\le$ 0.015 & J1203+0413 & 14.0 &           77  &  88 &  $\le$ 2.34&  $\le$ 2.5 & 3.8  \\
       3 & J0958+4725 &      763 & 0.076 & 1.85 & 4.4 &  $\le$ 0.017 & J0958+4719 & 13.2 &           87  &  7 & 0.16 $\pm$ 0.02& 1.1 & 0.8  \\
       4 & J1257+3229 &      589 & 0.079 & 1.85 & 4.3 &  $\le$ 0.022 & J1257+3228 & 17.4 &           47  &  101 &  $\le$ 17.54&  $\le$ 7.3 & 4.2  \\
       5 & J1228+3706 &      382 & 0.070 & 1.51 & 5.5 &  $\le$ 0.043 & J1228+3714 & 11.0 & 0$\rm ^d$  &  6 &  $\le$ 0.02&  $\le$ 0.2 & 0.8  \\
       6 & J1228+3706$^c$   &      382 &  $-$  &  $-$   &  $-$   &  $-$   & J1228+3706 & 15.2 & 0$\rm ^d$  &  653 &  $-$   &  $-$  & 9.6  \\
       7 & J0849+5108 &      344 & 0.088 & 1.85 & 4.2 &  $\le$ 0.037 & J0849+5108 & 19.9 & 0$\rm ^d$  &  332 &  $\le$ 152.54&  $\le$ 23.4 & 15.7  \\
       8 & J1030+2359 &      336 & 0.076 & 1.64 & 6.0 &  $\le$ 0.054 & J1030+2357 & 10.0 & 0$\rm ^d$  &  19 & 0.80 $\pm$ 0.03& 2.5 & 1.3  \\
       9 & J1232+1809 &      307 & 0.076 & 1.63 & 4.4 &  $\le$ 0.043 & J1232+1814 & 18.6 & 0$\rm ^d$  &  10 & 0.23 $\pm$ 0.03& 1.3 & 1.1  \\
      10 & J1221+3051 &      291 & 0.075 & 1.86 & 2.6 &  $\le$ 0.027 & J1222+3053 & 12.0 & 0$\rm ^d$  &  8 &  $\le$ 0.04&  $\le$ 0.3 & 0.8  \\
      11 & J1159+4412 &      236 & 0.076 & 1.65 & 3.2 &  $\le$ 0.041 & J1158+4411 & 10.9 &           89  &  10 &  $\le$ 0.01&  $\le$ 0.1 & 1.1  \\
      12 & J1218+1738 &      179 & 0.076 & 1.65 & 3.1 &  $\le$ 0.053 & J1218+1743 & 13.9 & 0$\rm ^d$  &  10 &  $\le$ 0.16&  $\le$ 0.6 & 1.4  \\
      13 & J1254+0233 &      152 & 0.076 & 1.65 & 2.1 &  $\le$ 0.041 & J1254+0239 & 18.6 &           53  &  14 & 2.51 $\pm$ 0.01& 4.6 & 2.3  \\
      14 & J1321+3510 &      144 & 0.076 & 1.64 & 1.9 &  $\le$ 0.040 & J1322+3512 & 19.7 &           78  &  9 &  $\le$ 0.03&  $\le$ 0.2 & 0.4  \\
      15 & J1440+0531 &      114 & 0.076 & 1.63 & 2.4 &  $\le$ 0.063 & J1440+0531 & 4.6 &           86  &  22 & 0.16 $\pm$ 0.01& 1.1 & 1.1  \\
      16 & J1015+1637 &      110 & 0.084 & 1.86 & 2.2 &  $\le$ 0.060 & J1015+1637 & 19.9 &           39  &  243 &  $\le$ 23.72&  $\le$ 8.6 & 8.0  
\enddata
\tablenotetext{a}{At the resolution of 1~\kms.}
\tablenotetext{b}{Integrated optical depth binned over 1 \kms bins i.e $\sim$12-14 pixels. The limiting values represent 3~$\sigma$ limiting optical depth, which are also used in Figure~5 and 6. }
\tablenotetext{c}{Data corrupted by RFI. }
\tablenotetext{d}{Face-on orientation of the host galaxy.}
\end{deluxetable}
\clearpage
\end{landscape}

\end{document}